\documentclass[twocolumn]{aastex62}
\usepackage[figuresright]{rotating}
\usepackage{multirow}
\usepackage{array}
\pdfoutput=1 %for arXiv submission
\usepackage{amsmath,amstext}
\usepackage{CJK}
\usepackage[T1]{fontenc}
\usepackage[figure,figure*]{hypcap}
\usepackage{booktabs}
\usepackage{cleveref}
\usepackage{paralist}
\usepackage{natbib}
\usepackage{graphicx}
\usepackage{hyperref} %to use the \url or \href tag
\graphicspath{{fig/}}

\newcommand{\kms}{km\,$\rm s^{-1}$}
\newcommand{\masyr}{\,mas\,$\rm yr^{-1}$}

\newcommand {\Usun}{{U_{\!\odot}}}
\newcommand {\Wsun}{{W_{\!\odot}}}
\newcommand {\Vsun}{{V_{\!\odot}}}

\newcommand {\thj}[1]  {{\bf\textcolor{black}{#1}}}
\newcommand {\wf}[1]  {{\bf\textcolor{black}{#1}}}

\shorttitle{The Stellar "Snake" I: Structure and Properties}
\shortauthors{ Wang et al.}
\begin{document}
\begin{CJK*}{UTF8}{gbsn}

\title{The Stellar "Snake" I: Whole Structure and Properties}
\correspondingauthor{Hai-Jun Tian}
\email{hjtian@lamost.org}

\author{Fan Wang (王凡)}
\affil{China Three Gorges University, Yichang 443002, People's Republic of China.}
\affil{Center for Astronomy and Space Sciences, China Three Gorges University, Yichang 443002, People's Republic of China.}

\author[0000-0001-9289-0589]{Haijun Tian (田海俊)}
\affil{China Three Gorges University, Yichang 443002, People's Republic of China.}
\affil{Center for Astronomy and Space Sciences, China Three Gorges University, Yichang 443002, People's Republic of China.}

\author{Dan Qiu (邱丹)}
\affil{Key Laboratory of Space Astronomy and Technology, National Astronomical Observatories, Chinese Academy of Sciences, Beijing 100101, People's Republic of China.}
\affil{ University of Chinese Academy of Sciences, Beijing 100049, People's Republic of China.}
\affil{Center for Astronomy and Space Sciences, China Three Gorges University, Yichang 443002, People's Republic of China.}

\author{Qi Xu (许祺)}
\affil{China Three Gorges University, Yichang 443002, People's Republic of China.}
\affil{Center for Astronomy and Space Sciences, China Three Gorges University, Yichang 443002, People's Republic of China.}

\author{Min Fang (房敏)}
\affil{Purple Mountain Observatory, Chinese Academy of Sciences, Nanjing 210023, People's Republic of China.}

\author{Hao Tian (田浩)}
\affil{Key Laboratory of Space Astronomy and Technology, National Astronomical Observatories, Chinese Academy of Sciences, Beijing 100101, People's Republic of China.}
\author[0000-0003-3010-7661]{Li Di}
\affil{National Astronomical Observatories, Chinese Academy of Sciences, Beijing 100101, People's Republic of China.}%
\affil{ University of Chinese Academy of Sciences, Beijing 100049, People's Republic of China}
\affil{NAOC-UKZN Computational Astrophysics Center, University of KwaZulu-Natal, Durban 4000, South Africa.}

\author{Sarah A. Bird}
\affil{China Three Gorges University, Yichang 443002, People's Republic of China.}
\affil{Center for Astronomy and Space Sciences, China Three Gorges University, Yichang 443002, People's Republic of China.}

\author{Jianrong Shi (施建荣)}
\affil{National Astronomical Observatories, Chinese Academy of Sciences, Beijing 100101, Peopleʼs Republic of China.}%
\affil{ University of Chinese Academy of Sciences, Beijing 100049, People's Republic of China}

\author{Xiaoting Fu (符晓婷)}
\affil{The Kavli Institute for Astronomy and Astrophysics at Peking University, Beijing 100871, People's Republic of China.}

\author{Gaochao Liu (刘高潮)}
\affil{China Three Gorges University, Yichang 443002, People's Republic of China.}
\affil{Center for Astronomy and Space Sciences, China Three Gorges University, Yichang 443002, People's Republic of China.}

\author{Sheng Cui (崔盛)}
\affil{China Three Gorges University, Yichang 443002, People's Republic of China.}
\affil{Center for Astronomy and Space Sciences, China Three Gorges University, Yichang 443002, People's Republic of China.}
\author{Yong ZHANG (张勇)}
\affil{Nanjing Institute of Astronomical Optics \& Technology, National Astronomical Observatories, Chinese Academy of Sciences, Nanjing 210042, People's Republic of China.}

%==============================================================

\begin{abstract}
To complement our previous discovery of the young snake-like structure in the solar neighborhood and reveal the structure's full extent, we build two samples of stars within the Snake and its surrounding territory from {\tt Gaia EDR3}. With the friends-of-friends algorithm, we identify 2694 and 9615 Snake member candidates from the two samples. Thirteen open clusters are embedded in these member candidates. By combining the spectroscopic data from multiple surveys, we investigate the comprehensive properties of the candidates and find that they \thj{are very likely to} belong to one sizable structure, since most of the components are well bridged in their spatial distributions, and follow a single stellar population with an age of $30-40$\,Myr and solar metallicity. This sizable structure is best explained as hierarchically primordial, and probably formed from a filamentary giant molecular cloud with unique formation history in localized regions. To analyze the dynamics of the Snake, we divide the structure into five groups according to their tangential velocities; we find that the groups are expanding at a coherent rate ($\kappa_X\sim3.0\,\times10^{-2}\,\rm km\,s^{-1}\,pc^{-1}$) along the length of the structure ($X$-direction). \thj{The corresponding expansion age ($\tau\sim33$\,Myr) is highly consistent with the age of the Snake}. With over ten thousand member stars, the Snake is an ideal laboratory to study nearby coeval stellar formation, stellar physics, and environmental evolution over a large spatial extent.

\end{abstract}


%================================================================================================
\section{Introduction}
\label{sec:intro}
Filamentary structures have a long history of detection within Galactic giant molecular clouds (GMCs) \citep{Schneider1979ApJS, Su2015ApJ, Zucker2018ApJ, Soler2021}; clouds of such type are the stellar factories of galaxies \citep{Dobbs2013}. One scenario of the evolution of these filaments results in elongated stellar structures, e.g., the stellar "Snake" -- a young (only $30-40$\,Myr) snake-like structure in the solar neighborhood (around 300\,pc from the Sun) that was first reported by \citet[hereafter T20]{Tian(2020)}. Besides bright O and B members, this structure includes a number of intermediate-mass stars and low-mass pre-main sequence (PMS) stars extending hundreds of parsecs. The Snake members are consistent with a single stellar population in the color-absolute magnitude diagram (CMD) and a continuous mass function. Thus the Snake is an ideal laboratory to study nearby coeval stellar formation, stellar physics, and environmental evolution over a large spatially extended region.

The Snake communes alongside several stellar populations of widely different ages as well as spatial and kinematic extent, for instance, the Vela-Puppis complex \citep[hereafter, CG18, CG19a, CG19b]{Cantat-Gaudin(2018a),Cantat-Gaudin(2019a), Cantat-Gaudin(2019b)} and Orion complex \citep[hereafter, K18]{Kounkel(2018)}. The community includes tens of open clusters, associations, and groups; and many of these populations have been presented in the literature \citep[e.g., K18; CG18; CG19a,b;][]{deZeeuw(1999),Beccari(2018),Armstrong(2018),Zari2019,Armstrong(2020), Pang2021arXiv210607658P}. These populations are spatially not far from each other, and their average distance is around 300\,pc from the Sun \citep[CG18;][]{Franciosini(2018), ChenBQ2020}. In terms of age, they are young with ages $<$\,50\,Myr 
\citep[CG19b;][]{Zari2019}.

The age of the Vela-Puppis complex has been derived by several works in the literature %, the ages of the open clusters have been measured repeatedly, but the values have yet to converge within the literature
\citep{Dias(2002), Kharchenko(2013), Bossini(2019)}. The average age ($30-40$\,Myr) is quite similar with that of the Snake (T20). In addition, there are several younger clusters in this region, e.g., BH\,23 and $\gamma$\,Vel, with ages of $\sim$10\,Myr \citep{Bossini(2019),Kharchenko(2013)}, and an even younger (only $3-4$\,Myr old) massive binary system $\gamma^2$\,Velorum \citep{Jeffries(2009)}. CG19a proposed that stellar feedback and supernovae of massive stars in the 30\,Myr old clusters may have produced a central cavity and shell, i.e., the IRAS Vela Shell \citep[IVS,][]{Sahu(1992)}, and triggered a second burst of star formation (resulting in $\gamma^2$\,Velorum) $\sim$10\,Myr ago. In the Orion complex, most components have an age of 1$-$12\,Myr \citep{Hillenbrand1997,Fang2009,Fang2013,Fang2017,Fang2021,Dario2016,Kounkel2018}, except the open cluster ASCC\,20 which has an age of 21\,Myr \citep{Kos(2019)}. Moreover, the average radial velocity of the Snake ($\sim 25\pm5$\,\kms, T20) is larger than that of the known clusters in the Vela-Puppis complex ($\sim16\pm3$\,\kms, Table 2 of CG19a and Table 1 of \citet{Kovaleva(2020c)}), and that of the Orion complex ($<15$\,\kms, Figure 11 of K18).
As for the formation mechanism, the stellar bridges connecting the open clusters found by \citet[hereafter, BBJ20]{Beccari(2020)} clearly demonstrate that the snake-like (T20) and string-like \citep{Kounkel(2019)} structures are primordially and hierarchically formed from filamentary relics in a GMC. On the other hand, evidence such as the expansion of the structures found by K18, CG19b, and T20, and the corona of nearby star clusters discovered by \citet{Meingast(2020)} make the situation more complicated.

%NGC\,2451/2547, Trumpler\,10, UBC\,7, $\gamma$\,Vel, BH\,23
%including the five components (i.e., Orion A, B, C, D, λ Ori according to \citet{Kounkel(2018)}
%including two open clusters (Tian\,2 and NGC\,2232) and a long tail

In this work, we take a census of the populations forming the Snake in its surrounding territory with {\tt Gaia EDR3}. We use the coherent algorithm (i.e., friends-of-friends, FOF) as employed by T20 to search for the full extent of the Snake and uncover its complete structure. Meanwhile, we add data from multiple spectroscopic surveys, e.g., {\tt LAMOST} \citep{Cui2012RAA}, {\tt APOGEE} \citep{Majewski(2017)}, {\tt GALAH} \citep{Buder(2021)}, and {\tt RAVE} \citep{Steinmetz(2020b)}, to depict the comprehensive properties (e.g., kinematics, chemistry, and so on) of the Snake and explore clues to its formation and evolution. 

We organize the paper as follows. In Section 2, we describe our data and target selection, and then illustrate the Snake and its nearby structure from the up-to-date {\tt Gaia EDR3} in Section 3. By combining data from multiple spectroscopic surveys, we explore the properties of the populations composing the Snake in Section 4. Finally, we provide a discussion and our conclusions in Sections 5 and 6, respectively.

Throughout the paper, we adopt the solar motion $(\Usun, \Vsun, \Wsun)$\,=\,(9.58, 10.52, 7.01)\,\kms \citep{Tian(2015)} with respect to the local standard of rest (LSR), and the solar Galactocentric radius and vertical position $(R_{0}, Z_{0})$\,=\,(8.27, 0.0)\,kpc \citep{Sch-Ralph(2012)}. In the gnomonic projection coordinate system, $l^{*}$ is used to denote the Galactic longitude, such that, for example, $\mu_{\rm l^{*}}=\mu_{\rm l}\cos{b}$. The proper motion $(\mu_{\rm l^{*}},\mu_{\rm b})$ for each star is corrected for the solar peculiar motion in Galactic coordinates.

%============================================================================================================
\section{Data}
\label{sec:data}

\subsection{Data sources}
\label{}
As in T20, we use 5D phase-space information, i.e., $l$, $b$, $\mu_{l^*}$, $\mu_b$, and distances, to uncover the member candidates from {\tt Gaia EDR3}. We obtain radial velocities and precise metallicities for the member candidates from the spectroscopic surveys {\tt LAMOST DR8}, {\tt GALAH DR3}, {\tt APOGEE DR14}, and {\tt RAVE DR6}.  We use the {\tt CO} survey \citep{Dame2001} to investigate the relationship between the Snake and its surrounding gas environment. In the following, we will briefly introduce these data sources.

\subsubsection{Gaia EDR3}
\label{}
{\tt Gaia EDR3} \citep{GaiaCollaboration(2021)} has provided celestial positions ($l$,\,$b$), parallaxes $(\omega)$, and proper motions ($\mu_l$,\,$\mu_b$) with unprecedented precision for more than 1.5 billion sources brighter than 21\,mag in the $G$ band, and has provided the magnitude in $G$, $G_\mathrm{BP}$, and $G_\mathrm{RP}$ with typical uncertainties of 0.2$-$6.0\,mmag for sources brighter than 20\,mag. {\tt Gaia EDR3} inherits median radial velocities ($\rm R_V$) from its previous data release ({\tt DR2}) for about 7.21 million bright sources ($4<G/\mathrm{mag}<13$). The overall precision of the $\rm R_V$ at the bright end is 0.2$-$0.3\,\kms\ while at the faint end, the overall precision is about 1.2$-$3.5\,\kms. Relative to {\tt Gaia DR2}, the precision has been increased by 30\% for parallaxes and by a factor of 2 for proper motions in {\tt Gaia EDR3}, while the systematic errors have been suppressed by 30\% to 40\% for the parallaxes and by a factor of $\sim$2.5 for the proper motions in {\tt Gaia EDR3}.

\subsubsection{LAMOST DR8}
\label{}

 {\tt LAMOST DR8} \citep{Zhao2012RAA,LiuC2020} includes spectra obtained from the low- and medium-resolution surveys (LRS and MRS). The LRS and MRS datasets provide the stellar spectral parameters for about 6.48 million and 1.24 million stars, respectively. The uncertainties are about 0.081\,dex (0.025\,dex) in [Fe/H] and 5.48\,\kms (0.9\,\kms) in $\rm R_V$ for the LRS (MRS) dataset. We add a correction to the LRS radial velocities which have a systematic underestimation of $\sim 5.7$\,\kms\citep{Tian(2015)}.
 
 %of 1.24 million stars are released in the dataset of MRS, in which the uncertainties is about 0.025\,dex for [Fe/H] and about 0.9\,\kms\ for {\tt RV}, and the stellar spectral parameters of 6.48 million stars are released in the LRS, which also include [Fe/H] with the uncertainty of 0.0081\,dex and {\tt RV} with about 5.48\,\kms. .
 %High-quality spectra with the signal-to-noise ratio over 10 reach the number of 13.28 million in the dataset of LRS.

\subsubsection{GALAH DR3}
\label{}
 {\tt GALAH DR3} \citep{Buder(2021)} obtained stellar parameters and elemental abundances for 678,423 spectra of 588,571 nearby stars\,(81.2\% of the stars are within\,$\sim$2\,kpc). For stars with $S/N>40$, the accuracy and precision are 0.034\,dex and 0.055\,dex in [Fe/H], 0.1\,\kms\ and 0.34\,\kms\ in $\rm R_V$, and 2.0 and 0.83\,\kms\ in rotational velocity.

\subsubsection{APOGEE DR14}
\label{}
{\tt APOGEE DR14} \citep{Majewski(2017)} has collected a half million infrared (1.51$-$1.70 $\mu$m) spectra with high-resolution ($R\sim22500$) and high $S/N$ ($>100$) for 146,000 stars, and time series information via repeated visits for most of these stars. The average precision is $\sim$0.1\,\kms\ in radial velocity, and $<0.1$\,dex for chemical abundances of about 15 chemical species \citep{Nidever2015AJ}.
%while  the  survey  provides external calibration sufficient to ensure accuracies at the level of\,$\sim$0.35\,\kms\.#0.018

\subsubsection{RAVE DR6}
\label{}
 {\tt RAVE} \citep{Steinmetz(2020b)} is a magnitude-limited ($9<I/\mathrm{mag}<12$) spectroscopic survey of randomly selected Galactic stars. {\tt RAVE} medium-resolution spectra (R$\sim$7500) span the region of the Ca-triplet (8410$-$8795\,\AA). Its sixth data release (DR6) contains stellar parameters for 416,365 stars with $S/N>20$. The typical uncertainties are $\sim$100\,K in $\rm T_{eff}$, 0.15\,dex in $\log g$, $<$2\,\kms\ in $\rm R_V$, and 0.10$-$0.15\,dex in metallicity. The typical $S/N$ of a star is 40.%$68\%$ of the sample has an internal accuracy better than 1.4\,km/s.

\subsubsection{CO map}
\label{}
\citet{Dame2001} released a {\tt CO} map of the entire Milky Way by combining 37 individual surveys (see their Table 1), which were conducted with the CfA 1.2\,m telescope and a similar instrument on Cerro Tololo in Chile. The composite {\tt CO} survey consists of 488,000 spectra covering the entire Galactic plane over a strip of $4\degr-10$\degr\ in latitude, and includes nearly all large local clouds at higher latitudes. The composite gas map helps to reveal large scale structures found in the molecular Galaxy.
%The composite map provides detailed information on individual molecular clouds, suggest relationships between clouds and regions widely separated on the sky, and clearly display the main structural features of the molecular Galaxy. In addition, the precise ($<2.0$\,\kms) kinematic information forms the foundation for many large-scale Galactic studies.

\subsection{Target selection}
\label{}
In this study, we mainly use the astrometric and photometric data from {\tt Gaia EDR3} \citep{GaiaCollaboration(2021)} to search for Snake member candidates. 

We define two search regions to investigate connections between the previously discovered neighboring snake-like (T20) and string-like (BBJ20) filemantary structures. According to T20 and BBJ20, we initially select two samples from {\tt Gaia EDR3}, i.e., Parts\,I and II, respectively, with the following empirical criteria:

1. $170^{\circ}<l<230^{\circ}$ and $-30^{\circ}<b<10^{\circ}$ for Part\,I, and $210^{\circ}<l<290^{\circ}$ and $-30^{\circ}<b<10^{\circ}$ for Part\,II, to ensure all members from T20 and BBJ20 are selected as corresponding members in our defined Parts\,I and II. We define an overlapping region included in both Parts\,I and II in order to search for member stars which bridge the two samples together.

2. $1.2<\omega/\mathrm{mas}<5.0$, and $\omega/\sigma_{\omega}>10.0$ for both Parts\,I and II, to restrict the samples in the effective volumes ($200<d/\mathrm{pc}<770$). Here, we derive distances by directly inverting parallax with $d=1000.0\omega^{-1}$ (pc).

3. $(\mu_{l^*}-\bar{\mu}_{l^*,m})^2+(\mu_{b}-\bar{\mu}_{b,m})^2<(5\sigma_{\mu,m})^2$, to restrict to stars with proper motions within $5\sigma_{\mu,m}$ of $(\bar{\mu}_{l*,m}, \bar{\mu}_{b,m})$. Here, $\bar{\mu}_{l^*,m}, \bar{\mu}_{b,m}$, and $\sigma_{\mu,m}$ are the average and root-mean-square ({\tt rms}) of the proper motions of the members in the space of $(\mu_{l^*},\mu_{b})$. According to T20 and BBJ20, we empirically set $(\bar{\mu}_{l^*,m}, \bar{\mu}_{b,m}) \simeq (-2.59,-2.0)$\,\masyr\ for Part\,I, $(-7.0,-2.0)$\,\masyr\ for Part\,II, and set $\sigma_{\mu,m} {\simeq 1.0}$\,\masyr\ for both Parts\,I and II.  

4. ${\tt RUWE} < 1.4$ for both Parts\,I and II, to limit to sources with high quality astrometric solutions. {\tt RUWE} is the renormalized unit weight error which is defined in \citet{Lindegren(2018)}.

These selection criteria yield 109,000 and 113,728 stars for Parts\,I and II, respectively. As in T20, the $V$-band extinction $A_{V}$ in magnitude is approximately $0.85\,\times\,d$. The extinctions in {\tt Gaia}'s bands for each star can be calculated from $A_{V}$ \citep{Tian(2014)}, using the extinction coefficients $(A_{\lambda}/A_{V})$ as 1.002, 0.589, and 0.789 for the $G_\mathrm{BP}$, $G_\mathrm{RP}$，$G$ bands, respectively.
%$\rm G_{BP}$, $\rm G_{RP}$, and $\rm G$ bands, respectively.
%$\mu$m

\section{Stellar Snake and its nearby structures}

\subsection{Membership}
\label{}

%Membership refinement by their concentration in multiple astrometric space is the most commonly used method for associations showing clear over densities in the astrometric space. 
As in T20, we adopt the FOF algorithm using the software {\tt ROCKSTAR} \citep{Behroozi(2013)} to search for members of the Snake from both Parts\,I and II. \wf{{\tt ROCKSTAR} employs a technique of adaptive hierarchical refinement in 6D phase space. It divides all the stars into several FOF groups by tracking the high number density clusters and excising stars that are not grouped in the star aggregates.} Using the 5D phase information (i.e., $l$, $b$, $\mu_{l^*}$, $\mu_b$, and distance) of each star as the input parameters \wf{(noting that $v_{\rm los}$ is set to zero for each star)}, the optimized {\tt ROCKSTAR} \citep{TianH_PHD} will automatically adjust the linking-space between members of "friend" stars, and divide them into several groups, simultaneously removing isolated individual stars from the groups. In this step, we find 3316 member candidates in Part\,I, and 9354 member candidates in Part\,II. % 9917 . 
This process successfully recovers the member stars of the open clusters that were previously defined as member clusters by T20 and BBJ20, i.e., Tian\,2, Trumpler\,10, Collinder\,135/140, NGC\,2232/2451B/2547, UBC\,7, and BBJ\,1/2/3. \thj{Two open clusters (Collinder\,132 and Haffner\,13) defined as members by BBJ20 are not recovered in this process.}

\wf{According to BBJ20, Collinder\,132 and Haffner\,13 are open clusters belonging to our defined Part\,II. To obtain the member candidate stars of Collinder\,132 and Haffner\,13, we separately run {\tt ROCKSTAR} on another sub-sample which is selected from {\tt Gaia EDR3} with similar criteria as Part\,II, but queried in a smaller region of the sky ($230^{\circ}<l<255^{\circ}$ and $-15^{\circ}<b<5^{\circ}$) and with different average proper motion, i.e., $(\bar{\mu}_{l^*,m},\bar{\mu}_{b,m})\simeq(-5.0,-0.4)$\,\masyr\,(BBJ20). In this step, Collinder\,132 and Haffner\,13 contribute 172 and 391 member candidates, respectively, for Part\,II. }

\wf{We remove the 430 duplicate member candidates from Part\,I in the overlapping bridge region between Parts\,I and II. Finally we retain 2886 and 9917 candidates member stars of Parts\,I and II, respectively.}

%The Haffner\,13 is an isolated cluster due to its larger distance than other sub-structures, it is why we can not well obtain the members of the Haffner 13 in this step. So we manually search for the member candidates for this cluster with the following criteria (xxx, 20xx): \\(1) 1.5\,<\,$\omega$\,<\,3.5;\\
%(2) $(\mu_{l^*}-\bar{\mu}_{l^*,m})^2+(\mu_{b}-\bar{\mu}_{b,m})^2<(5\sigma_{\mu,m})^2$.\\
%Here, we set $(\bar{\mu}_{l^*,m},\bar{\mu}_{b,m})\simeq(-6.6,-0.1)$\,\masyr\, and $\sigma_{\mu,m}\simeq0.3$\,\masyr. Then, we obtain 259 members candidates for the Haffner 13. Thus, we get 9342 (how do you get this number???) member candidates of Part\,II. 

We cross-match the 2886 and 9917 candidates using a radius of 1 arcsecond with the spectroscopic data from {\tt LAMOST MRS} \citep{LiuC2020}, {\tt LAMOST LRS} \citep{Zhao2012RAA}, {\tt APOGEE DR14} \citep{Majewski(2017)}, {\tt GALAH DR3} \citep{Buder(2021)}, and {\tt RAVE DR6} \citep{Steinmetz(2020b)} to obtain radial velocities, metallicities, rotational velocities, and other stellar parameters.

As in T20, we remove the possible contaminants whose ages are beyond the range between 5\,Myr and 120\,Myr, or those which obviously deviate from PMS or main sequence (MS) in the the CMD (see Figure \ref{fig:cmd_each} in Section \ref{sec:mass_age}). In this step, we eliminate 186 and 280 candidates from Parts\,I and II, respectively.

Additionally, we only keep candidates with radial velocities within 3$\sigma$ about the mean value. The outliers are either spurious members, or true members but with binarity, which are located on the binary sequence in the CMD. In this step, 6 and 22 stars are rejected from Parts\,I and II, respectively. Finally, we obtain 2694 and 9615 member candidates belonging to Parts\,I and II, respectively. 

%we find that the radial velocity with LSR of the candidates matched different catalogues has a difference with Gaia's radial velocity, and these large radial velocity stars with LSR are most likely binary candidates which are also located on the binary sequence in the CMD \citep[see, e.g.][for details]{Kouwenhoven(2008a)}, and then the memberships are further confirmed by their velocity with LSR, e.g., -10\,<\,$\upsilon$\,<\,30\,km/s, and then 23, 116, 17, 24 and 0 member candidates for part1 are matched with LAMOST MRS, LAMOST LRS, {\tt GALAH DR3}, APOGEE DR14 and RAVE DR6, respectively, and 0, 0, 45, 1 and 3 member candidates for Part\,II, respectively. Finally, the member candidates are 2849 and 8670 for Part\,I and Part\,II, respectively. 

%----
\begin{figure*}[!htbp]
  \centering
  \includegraphics[width=1.1\textwidth, trim=1.8cm 0.0cm 0.0cm 0.0cm, clip]{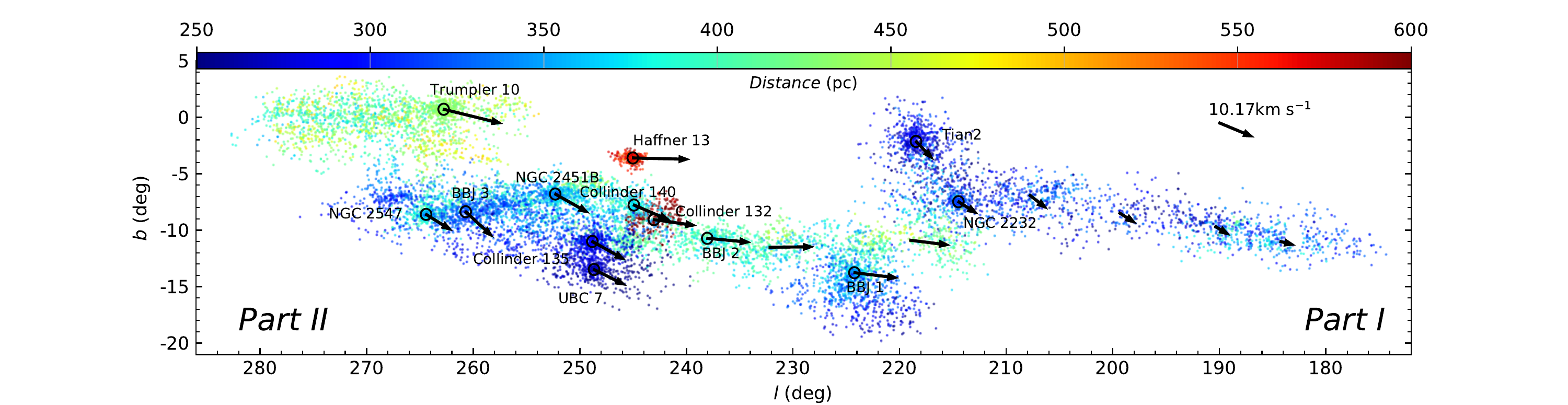}
  \caption{Spatial distribution ($l$-$b$) of the member candidates (color-coded with distance as represented in the color bar). The arrows over structures illustrate the average tangential velocities of stars in localized regions, while the arrow in the top-right corner marks the mean tangential velocity ($v_{l}= -9.47$\,\kms, $v_{b}= -3.72$\,\kms) of the whole structure. \wf{The lengths of the arrows are scaled with respect to the total tangential velocity of the whole structure (Parts\,I and II). All velocities are given with respect to the LSR.}}
  \label{fig:lb_pm}
\end{figure*}
%----

\subsection{Spatial Distribution}
\label{}
Figure \ref{fig:lb_pm} displays the distribution of the member candidates in the $l-b$ projected space. Each candidate is color-coded by distance. The black arrow in the top-right corner illustrates the average tangential velocity ($v_{l}=-9.47$\,\kms, $v_{b}=-3.72$\,\kms) of the whole sample. The remaining arrows display the average tangential velocities of localized regions. The arrow length corresponds to the value of velocity. 

\begin{figure*}[!htbp]
  \centering
  \includegraphics[width=1\textwidth, trim=0.1cm 0.1cm 0.0cm 0.0cm,clip]{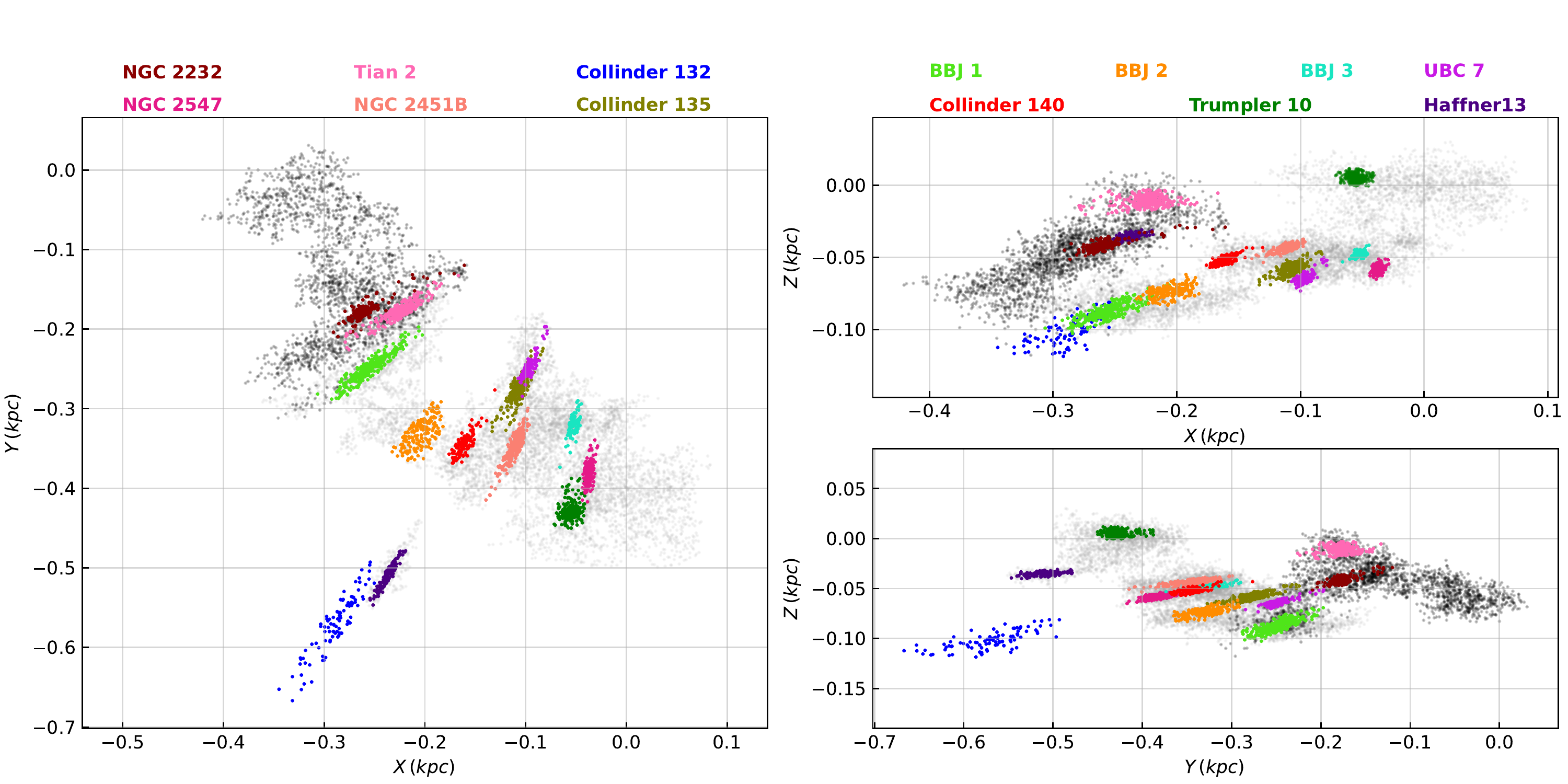}%0907
  \caption{Spatial distribution of the member candidates in Cartesian coordinates $(X, Y, Z)$. The candidates in Parts\,I and II are shown with the black and gray dots, respectively. The 13 open clusters are color-coded as specified in the legend located above the figure. The coordinates of the Sun are $(X, Y, Z) = (0.0, 0.0, 0.0)$\,kpc.
  }
  \label{fig:xyz}
\end{figure*}

Figure \ref{fig:xyz} displays the spatial distribution of the member candidates in the Cartesian coordinates $(X, Y, Z)$ for Part\,I (black dots) and Part\,II (gray dots), respectively. Both the length (in the $X$-direction ) and width (in the $Y$-direction) are around 500\,pc, but the thickness (in the $Z$-direction) is just over 100\,pc.

As shown in Figure \ref{fig:lb_pm} and \ref{fig:xyz}, Parts\,I and II are well bridged \thj{except for Haffner\,13 and Collinder\,132. These two distant clusters are also not well bridged with the main structure as presented in BBJ20 (see their Figure 8).} %This suggests that Parts\,I and II are very likely to be a single large-scale structure, as proposed by T20. This will be further demonstrated in Section \ref{sec:property}.% \wf{and discuss that Haffner\,13 and Collinder\,132 have a possible to belong to Parts\,I and II in details in Section \ref{sec:origin}.} %Therefore, all the substructures in both Parts\,I and II are components of the Snake.

%Both figures directly suggest that Part\,I and Part\,II are closer to each other. The 3D morphology of Part\,I and Part\,II again confirms the presence of the 2D elongation that we observed.

%Figure 1 shows the member distributions in Galactic space. The black and grey points in the space of l-b  clearly demonstrate the shape of Part\,II and Part\,II, and the two parts are closely connected. The tangential velocities of the clusters in the region are shown as black arrows with Local Standard of Rest (LSR). The mean proper motion of the whole stellar population is $\mu_{l^*}$= -8.95\,km/s, $\mu_{b}$= -3.75\,km/s (black arrow in the upper-left panel of Figure\,1). As shown in Figure\,2, both the length (in the x-direction ) and width ( in the y-direction) are around 500 pc, but the thickness ( in the z-direction ) is over 120\,{\tt pc}. Both figures directly suggest that Part\,I and Part\,II are closer to each other. The 3D morphology of Part\,I and Part\,II again confirms the presence of the 2D elongation that we observed.

%----
%----

\subsection{Open clusters}
\label{sec:open_clusters}

We identify 13 known open clusters in total (as shown in Figures \ref{fig:lb_pm}$-$\ref{fig:vlb}). Part\,I includes NGC\,2232 and Tian\,2 (T20), and Part\,II includes Trumpler\,10, Haffner\,13, NGC\,2547, NGC\,2451B, Collinder\,132, Collinder\,135, Collinder\,140, UBC\,7, BBJ\,1, BBJ\,2, and BBJ\,3 (BBJ20). These clusters are displayed with different colors in Figures \ref{fig:xyz} and \ref{fig:vlb}. In Figure \ref{fig:xyz}, the distribution of the clusters clearly shows a substantial elongation effect along the line of sight. This is induced by the deduced distances from inverse parallaxes (i.e., $d=1000\omega^{-1}$ (pc)). Comparing with BBJ20, we find finer sub-structures in our Part\,II region. For example, the clearly distinguishable physical pair of open clusters \citep{Kovaleva(2020c)}, i.e., Collinder\,135 and UBC\,7 (see Figure \ref{fig:lb_pm}). In addition to this pair, Tian\,2 and NGC\,2232 show characteristics of a physical pair according to their physical parameters, which will be further discussed in Section \ref{sec:property}. 

%{\bf How did you select the members for each open cluster? and how many members does each cluster has? please give a paragraph to describe this.}

%In order to explicitly display the 13 open clusters, we select 13 sub-samples according to their Galactic coordinates, e.g., $(l-l_0)^2+(b-b_0)^2 < {\epsilon_0}^2$, $l_0$, $b_0$, and $\epsilon_0$ are the coordinate of the center of the core, and the angular radius of the cluster. We roughly estimated $(l_0, b_0, \epsilon_0)=(214.4^\circ, -7.5^\circ, 2^\circ)$ for NGC\,2232, $(218.3^\circ, -2.1^\circ, 1.5^\circ)$ for Tian\,2, $(262.7^\circ, 0.7^\circ, 2^\circ)$ for Trumpler\,10, $(245.1^\circ, -3.6^\circ, 0.5^\circ)$ for Haffner\,13, $(264.4^\circ, -8.6^\circ, 1.0^\circ)$ for NGC\,2547, $(252.3^\circ, -6.8^\circ, 1.0^\circ)$ for NGC\,2451B, $(242.9^\circ, -1.1^\circ, 2.0^\circ)$ for Collinder\,132, $(248.8^\circ, -11.0^\circ, 1.5^\circ)$ for Collinder\,135, $(244.9^\circ, -7.7^\circ, 1.0^\circ)$ for Collonder\,140, $(248.7^\circ, -13.5^\circ, 1.5^\circ)$ for UBC\,7, $(224.2^\circ, -13.7\circ, 1.5^\circ)$ for BBJ\,1, $(238.2^\circ, -10.7^\circ, 2.0^\circ)$ for BBJ\,2, $(260.7^\circ, -8.3^\circ, 1.0^\circ)$ for BBJ\,3. According to this, 

We roughly select the member candidates from the sample in Parts\,I and II for the 13 open clusters according to their coordinates, i.e., $(l-l_0)^2+(b-b_0)^2 < {\epsilon_0}^2$. The central coordinate ($l_0$, $b_0$) is specified in Table \ref{tab:sums} for each cluster, and the angular radius ($\epsilon_0$) is empirically chosen to be 1.0\degr$-$2.0\degr\ for each individual cluster. Thus, we obtain 304, 213, 249, 148, 291, 260, 104, 282, 137, 183, 344, 184, and 88 member candidates for NGC\,2232, Tian\,2, Trumpler\,10, Haffner\,13, NGC\,2547, NGC\,2451B, Collinder\,132, Collinder\,135, Collonder\,140, UBC\,7, BBJ\,1, BBJ\,2, and BBJ\,3, respectively.

In Figure \ref{fig:vlb}, the open clusters present different groups in the tangential velocity space ($v_{l}-v_{b}$). According to the patterns in velocity, we define five groups, namely, Trumpler\,10 and its surrounding stars as Group\,I,  Collinder\,132, BBJ\,1, BBJ\,2, and their surrounding stars as Group\,II, NGC\,2547, NGC\,2451B, BBJ\,3, Collinder\,135, UBC\,7, Collinder\,140, and their surrounding stars as Group\,III, and Tian\,2, NGC\,2232, and its long tail as Group\,IV. Haffner\,13 is relatively isolated in both the 3D spatial distribution (the indigo dots in Figure \ref{fig:xyz}) and the tangential velocity distribution (the indigo dots in Figure \ref{fig:vlb}). Therefore, Haffner\,13 and its surrounding stars are labeled as Group\,V. We will investigate the 3D expansion rate for each group in Section \ref{sec:3d_ex_r}.
%it is probably in the same population as other clusters, because it has similar distribution in the CMD.
%According to this, we divide them into four groups to investigate their 3D expansion rates in Section \ref{sec:3d_ex_r}.

% the Trumpler\,10 has the largest tangential velocity

%To compare the twelve open clusters with the whole structure, we select the member candidates for these clusters according to their coordinates. We roughly estimate for these clusters, and Figure\,1 show Galactic coordinates and the distances for each panel. The symbol sizes present the distances derived by inverting parallax with $d = 1000.0/\omega$\,{\tt pc} and the different open clusters can be compared with each other. The darker the color, the greater the density. Figure\,3 separately display  distributions of the 2D velocities of the twelve clusters in same unit. The distribution of the radial velocities look not good, because there are a few available radial velocities for the 12 clusters. Besides, due to lack of metallicities for some clusters, we can't be sure to the metallicities for these clusters.  The ages of the 12 clusters are about 33.8\,{\tt Myr} to 35\,{\tt Myr} derived by T20 and B20, respectively, and we can also see their Galactic spatial distribution. 

%============================================================================================================
\section{Population Properties}
\label{sec:property}

We have 2694 and 9615 member candidates which can be used to analyze the statistical properties for Parts\,I and II, respectively.

\subsection{Kinematics}
\label{sec:motions}

\begin{figure}[!htbp]
  \centering
  \includegraphics[width=0.45\textwidth]{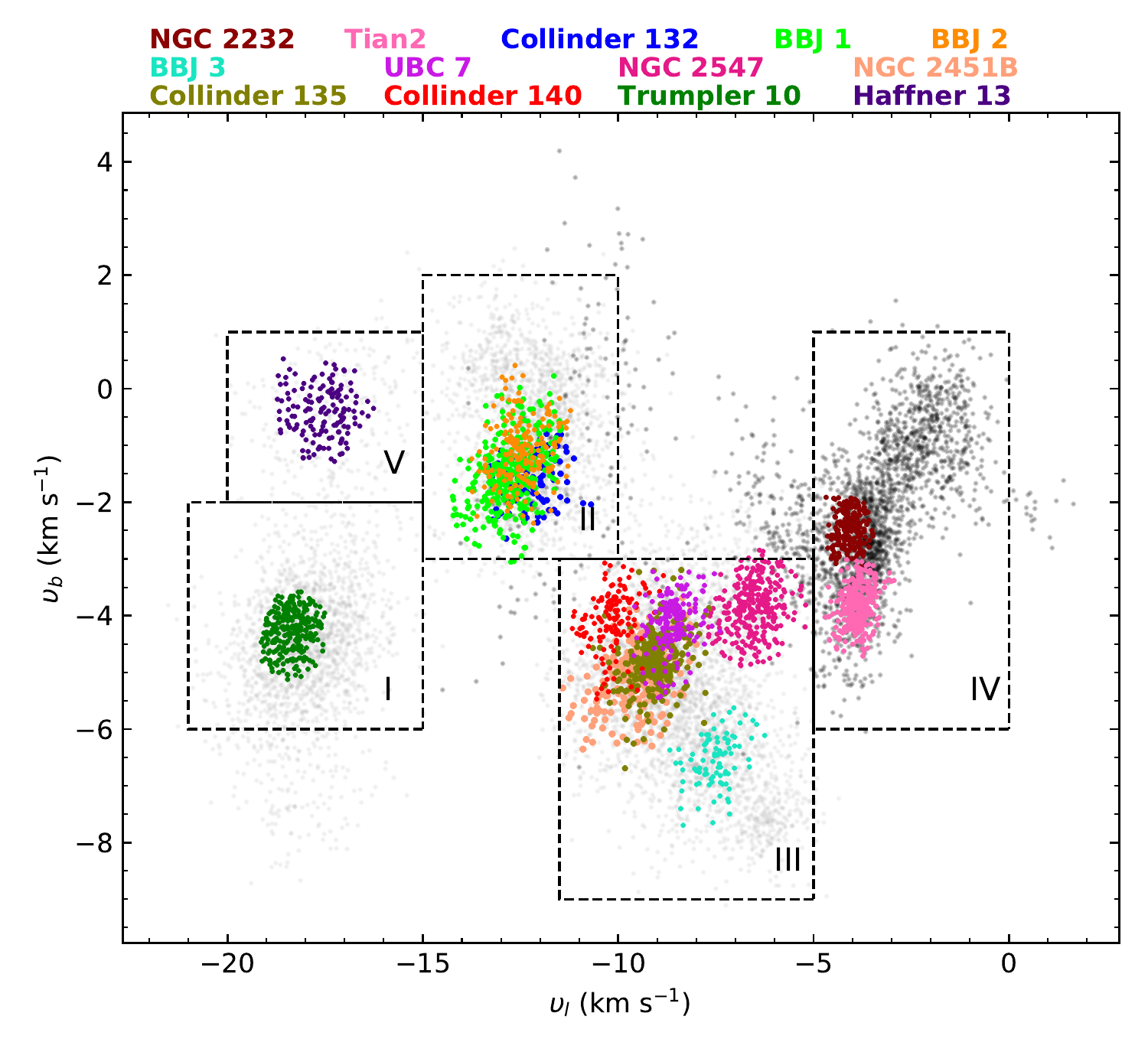}
  \caption{Distribution of tangential motions ($v_{l}-v_{b}$) for member stars of Parts\,I (black) and II (gray). The 13 open clusters are characterized by different patterns in velocity. Accordingly, we divide our sample into five groups. All velocities are with respect to the LSR. The color-coding is the same as in Figure \ref{fig:xyz}.
  }
  \label{fig:vlb}
\end{figure}

Over half of the identified open clusters are clearly distinguishable in $v_{l}-v_{b}$, as shown in Figure \ref{fig:vlb}, except for BBJ\,1, BBJ\,2, and Collinder\,132 in Group\,II and Collinder\,140 and Collinder\,135 in Group\,III. Notably, the two physical pairs, Collinder\,135 (olive) and UBC\,7 (magenta) in Group\,III and Tian\,2 (hot pink) and NGC\,2232 (dark red) in Group\,IV, are clearly separated in tangential velocity space.

%The Group\,I, mainly includes Trumpler\,10, has the largest tangential velocities. The clusters in the Group\,II are hard to be separated.

%For Part\,I, the average proper motion $({\mu}_{l^*},\mu_{b})$ is about $(-3.62\pm0.91, -2.51\pm1.37)$\,km/s, as shown by the black points in the Figure\,4; for Part\,II, the average proper motion $(\mu_{l^*},\mu_{b})$ is about $(-10.08\pm3.33, -4.21\pm2.50)$\,km/s shown by the grey points in the Figure\,3. As shown in the Figure\,3, we convert $\mu_{l^*}$ and $\mu_{b}$ into $v_{l}$ and $v_{b}$, and please note that $\mu_{l^*}$ and $\mu_{b}$ were estimated as $v_{l}$=4.74$\times\mu_{l^*}$/parallax and $v_{b}$=4.74$\times\mu_{b}$/parallax. Such velocities are better suited to search for kinematically clustered structures in the observed region than observed proper motions, since the spatial extent of the two parts is a significant fraction of the distance to it. Thus, converting to velocities resolves the distance degeneracy affecting observed proper motions in (mas/yr).

%----
%----
%$(-3.62\pm0.91, -2.51\pm 1.37)$\,\kms\, $(-10.08\pm3.33, -4.21\pm2.50)$\,\kms\ for the part1, the part2

%In order to verify that the group of stars discovered in proper motion space (see Figure\,1) is kinematically bound, we derived the radial (line-of-sight) velocities. 
Among the candidates in Part\,I, only 272 stars have radial velocities, of which 111 stars from {\tt Gaia}, 23 stars from {\tt APOGEE DR14}, 22 stars from {\tt LAMOST MRS} (11 stars with {\tt Gaia}'s $R_V$) and 104 stars from {\tt LAMOST LRS} (38 stars with {\tt Gaia}'s $R_V$), 12 stars from {\tt GALAH DR3} (7 stars with {\tt Gaia}'s $R_V$). Among the candidates in Part\,II, only 734 stars have radial velocities, of which 681 stars from {\tt Gaia}, 3 stars from {\tt RAVE} (2 stars with {\tt Gaia}'s $R_V$), 48 stars from {\tt GALAH DR3} (30 stars with {\tt Gaia}'s $R_V$) and 2 stars from {\tt APOGEE DR14}. For those stars that have been observed by multiple surveys,  we use the median values as their observed radial velocities.  

\begin{figure*}[!htbp]
  \centering
  \includegraphics[width=1\textwidth]{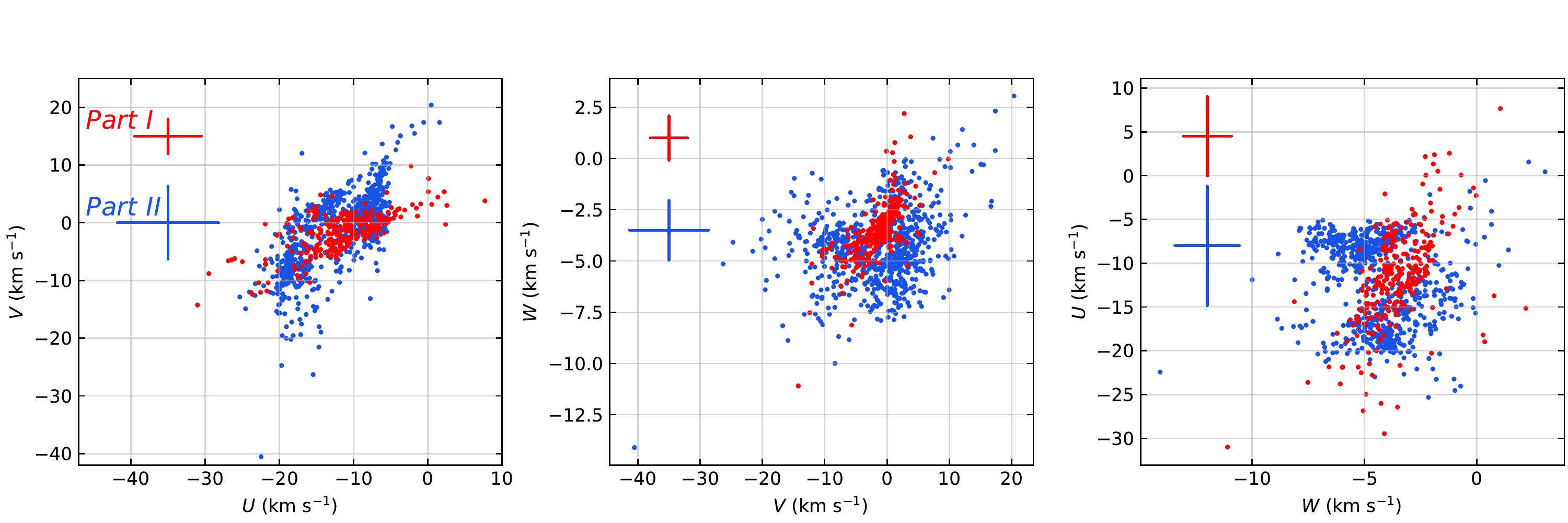}
  \caption{Distributions of the 3D $(U, V, W)$ velocities of the member candidates with $R_V$ in Part\,I (red) and Part\,II (blue). The 3D components $(U, V, W)$ for each candidate have been corrected for the solar peculiar velocity. The mean and {\tt rms} of the velocity components are  $(-11.18\pm4.60, -1.07\pm3.04, -3.62\pm1.10)$\,\kms\ for Part\,I and $(-11.05\pm6.88, -0.35\pm6.41, -4.42\pm1.47)$\,\kms\ for Part\,II. The error bars in the top-left corner of each panel mark the {\tt rms} of the 3D velocity components for Part\,I (red) and Part\,II (blue).}
  \label{fig:uvw}
\end{figure*}

Figure \ref{fig:uvw} displays the 3D ($U,V,W$) velocity distributions for Part\,I (red) and Part\,II (blue) in the Cartesian coordinate, mainly for the four groups, i.e., Group\,I$-$IV (Group\,V only has 7 radial velocity available). \thj{The mean and {\tt rms} of velocity components $(U,V,W)$ are  $(-11.18\pm4.60, -1.07\pm3.04, -3.62\pm1.10)$\,\kms\ for Part\,I and $(-11.05\pm6.88, -0.35\pm6.41, -4.42\pm1.47)$\,\kms\ for Part\,II. It indicates that the 3D velocity of Part\,I is highly consistent with that of Part\,II.} The outliers in the three sub-panels are likely peculiar members that have unresolved partners or high rotational speeds.

%----
%----

\begin{figure*}[!htbp]
  \centering
  \includegraphics[width=1\textwidth]{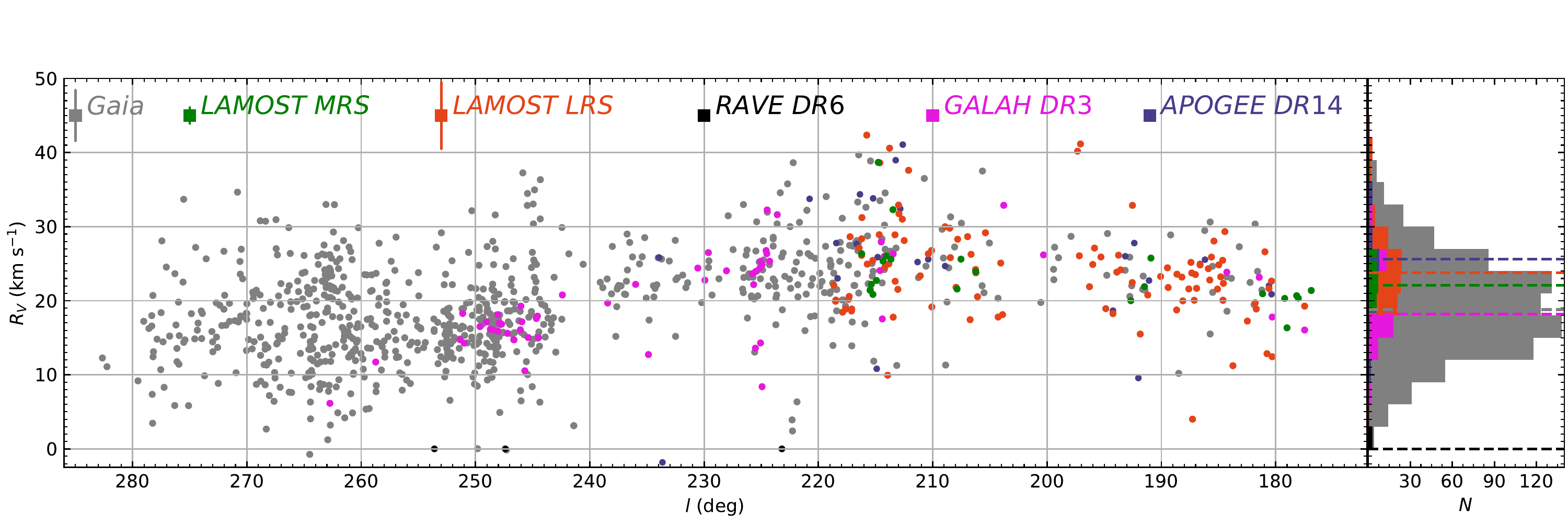}
  \caption{Distribution of the observed $R_V$ for all member candidates along Galactic longitude $l$. The $R_V$ are obtained from the multiple surveys (color-coded as specified in the legend). The error bars on each marker in the legend (top of the main panel) represent the average uncertainties of $R_V$ from the different surveys. The histograms of $R_V$ are displayed in the right sub-panel. The average values are marked with dashed lines. The color-coding is the same as the main panel.}
  \label{fig:lrv}
\end{figure*}

\begin{figure}[!htbp]
  \centering
  \includegraphics[width=0.45\textwidth]{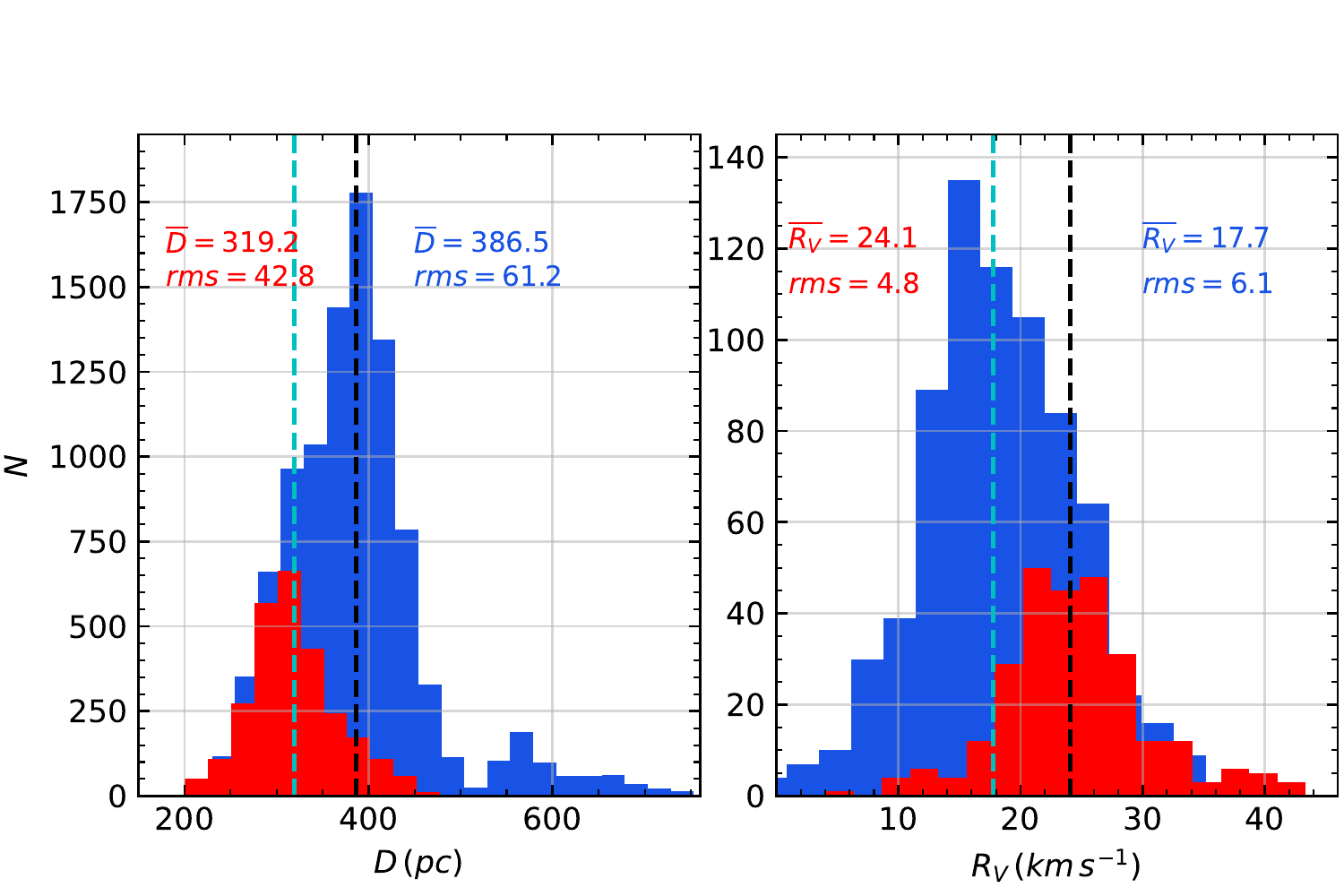}%0908
  \caption{Histograms of distance from the Sun (left panel) and observed radial velocity (right panel) for member candidates from Part\,I (red) and Part\,II (blue). The minor peak at $d\sim570$\,pc in the left panel is mainly due to the presence of Haffner\,13 and Collinder\,132.
  }
  \label{fig:drv}
\end{figure}

In Figure \ref{fig:lrv}, we show the variation of radial velocity $R_V$ along Galactic longitude $l$ for both Parts\,I and II. The radial velocities from different surveys are color-coded. For Part\,I, the radial velocities do not significantly change with Galactic longitude. For Part\,II, however, it seems that the radial velocities are inversely proportional to Galactic longitude. The average $R_V$ with uncertainty is $24.8\pm0.3$\,\kms\ for Part\,I, which is systematically about $7$\,\kms\ faster than that in Part\,II ($17.7\,\pm0.2$\,\kms), as shown in the histogram of $R_V$ in the right panel of Figure \ref{fig:drv} (red for Part\,I, and blue for Part\,II). The distance from the Sun of the member candidates in Part\,II are on average 67\,pc more distant than Part\,I (left panel of Figure \ref{fig:drv}).

%Interestingly, it seems that the candidates in Part\,II includes some members of Part\,I, since the histogram of the radial velocities of Part\,II has a sub-peak which matches with the distribution of radial velocities of Part\,I. This phenomenon looks more significant in the histograms of the distances in the left sub-panel of Figure\,\ref{fig:drv} ({\bf Please check again here, the duplicated sources in Part\,II have been removed from Part\,I}). It suggests that Parts\,I and II are difficult to be separated from each other, they should be a whole structure, even through the the average distance in Part\,II is systematically larger $\sim60$\,pc than that in Part\,I. 

We summarize the average velocities and distances for the 13 open clusters in Table \ref{tab:sums}.

%----
%----

%----
%----

\subsection{Chemistry and Surface Rotation}
\label{sec:chemistry}

\begin{figure*}[!htbp]
  \centering
  \includegraphics[width=1\textwidth]{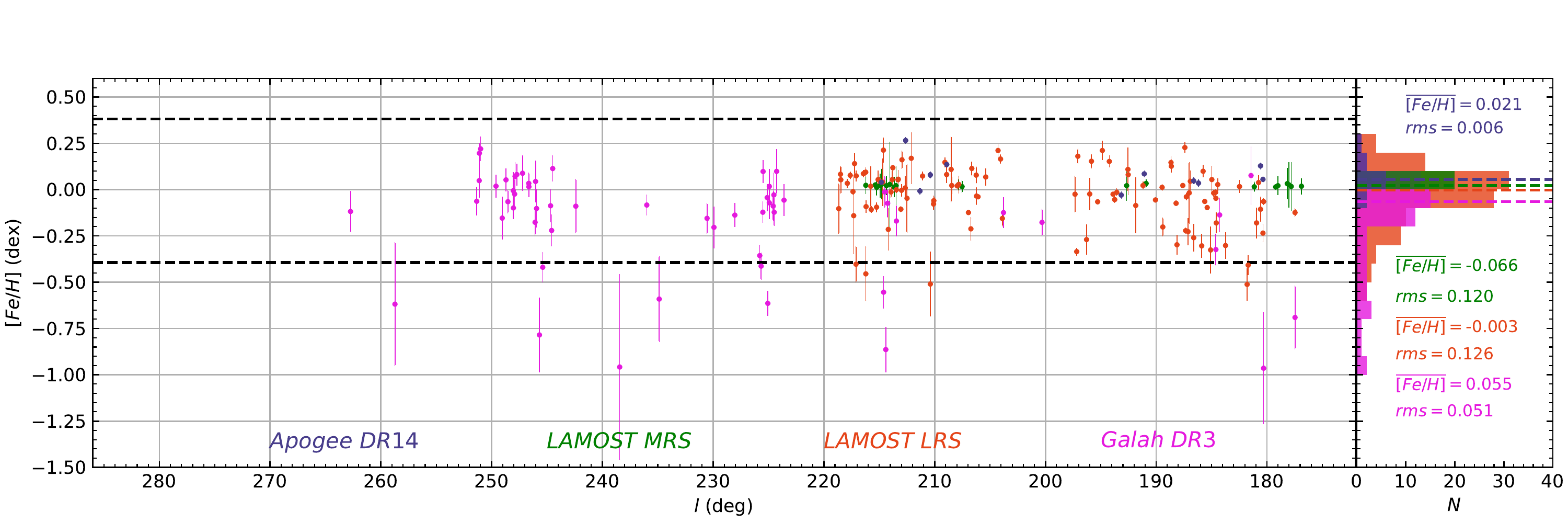}
  \caption{Distribution of [Fe/H] along Galactic longitude $l$. The values of [Fe/H] are obtained from multiple catalogs (color-coded as specified in the legend). The average and {\tt rms} of [Fe/H] are $-0.024\pm0.130$\,dex for all 202 available candidates. The two black dashed horizontal lines mark the locations of $\pm3\sigma$ around the average metallicity in the left main panel. The histograms of [Fe/H] from the different surveys are displayed in the right sub-panel. The average values are marked with dashed lines, and specified with the color-coded text. The color-coding is the same as Figure \ref{fig:lrv}.
  }
  \label{fig:metallicity}
\end{figure*}

Figure \ref{fig:metallicity} displays the variation of metallicity [Fe/H] (taken from various surveys) along Galactic longitude $l$ for member candidates. Specifically, we show 20 stars (red) from {\tt LAMOST MRS}, 105 stars (green) from {\tt LAMOST LRS}, 11 stars (dark slate blue) from {\tt Apogee DR14}, and 58 stars (magenta) from {\tt Galah DR3}. As shown in the figure, [Fe/H] spans a range of values between $-1.0$\,dex and $0.25$\,dex. The sample average from each survey is near solar ([Fe/H]$\sim0$ dex), as shown by the histograms in the right sub-panel. Considering that the Snake is much younger than our Sun, we prefer to use $Z=0.02$ (slightly larger than the solar value of $Z_{\odot}=0.015$) in the following discussion.

The average and {\tt rms} [Fe/H] for the whole sample are $-0.024$\,dex and $0.130$\,dex, respectively. The two black dashed horizontal lines mark the locations of $\pm\,3\sigma$. We consider the points beyond the two dashed lines as outliers. All the outliers are more metal poor than the mean metallicity of the Snake, particularly between $250\degr<l<245\degr$ where there seems to be a dip in Figure \ref{fig:metallicity}.

\begin{figure}[!htbp]
  \centering
  \includegraphics[width=0.45\textwidth]{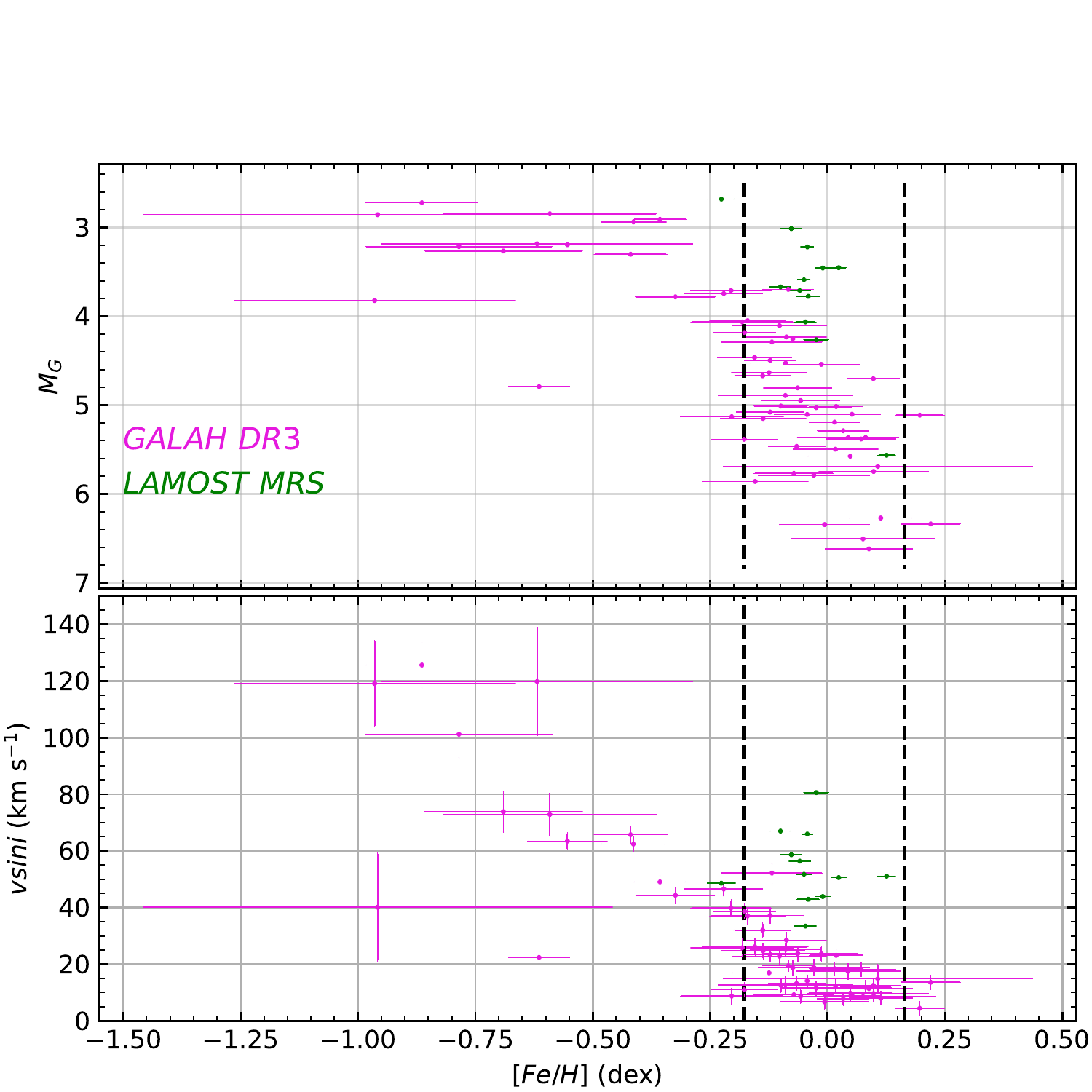}
  \caption{Comparison of [Fe/H] with absolute magnitude $M_G$ (top panel) and surface rotational velocity {\tt vsini} (bottom panel) from {\tt LAMOST MRS} (green) and {\tt Galah DR3} (magenta). The black dashed vertical lines mark the locations of $\pm3\sigma$ around the average metallicity ($\sigma\simeq0.13$\,dex, same as that in Figure \ref{fig:metallicity}).
  }
  \label{fig:rotation}
\end{figure}

We investigate the relationship between [Fe/H] and absolute magnitude $M_G$ (top panel) and surface rotation $v\sin{i}$ (bottom panel) of member candidates in Figure \ref{fig:rotation}. We find that all outliers in [Fe/H] are bright stars ($M_G<3.5$\,mag) or those with high surface rotation ($v\sin{i}>50$\,\kms). This demonstrates that accurate metallicity measurements may be hampered by high stellar temperature or fast rotation.

Based on these results, we will use this data set to more thoroughly study stellar physics and the possible uncertainties involved in measuring stellar parameters such as surface rotation, effective temperature, and metallicity (Wang Fan et al. in preparation).

%----

%----
%----

\subsection{Age and Mass}
\label{sec:mass_age}
%Stellar masses are sampled from the \citet{Kroupa(2001)} initial mass function (IMF). In Figure\,9, we show the mass function curves of Part\,I and Part\,II. And The change trend of the two curves is basically the same. However, when $log(M/M_{\bigot})$ is around 0.3, and the trends of the two curves are a little different, which Part\,I rises and descends and Part\,II always descends. And intuitively speaking, the number of stars first decreases and then increases slightly around 0.3, but we don't know what the real reason is, and we will study it further in later works.(This is will be ensured.)

%Isochrone fitting is a typical method to estimate stellar ages. \citet{Liu(2019)} provides a reliable implementation for this method. First, we prepare a series of isochrones from the Padova database Marigo(2017) with ages ranging from $log(\tau/yr)$\,=\,6.6 to 10.13 with an interval of $\Delta log(t/yr)$\,=\,0.01, and metallicities from $log(Z/Z_{\bigodot})$\,=\,-2.0 to 0.5 with an interval of 0.25. Then, we feed the colour index $G_{BP}-G_{RP}$, and the absolute magnitude $M_{G}$ of the member stars to the pipeline. Finally, the best fit isochrone can be found by minimizing the distance (defined as Equation\,2 in \citet{Liu(2019)}) between an isochrone and member stars. For simplicity, we assume a solar metal fraction of Z\,=\,0.0152.

T20 measured the age of the Snake (formally only Part\,I) using two different approaches: one based on isochrone fitting \citep{Liu(2019)}; and another using the Turn-On \citep{Cignoni2010ApJ}. Both methods give an age of 30$-$40\,Myr for the Snake. \citet{Binks(2021)} measured an age of 38$\pm$3\,Myr through the lithium depletion boundary for the open cluster NGC\,2232, which is one of the most important components of the Snake. Therefore, we firmly believe that the Snake is 30$-$40\,Myr old. In the following, we use an age of 34\,Myr, which is the best fit value in T20, for both Parts\,I and II.

\begin{figure}[!htbp]
  \centering
  \includegraphics[width=0.45\textwidth]{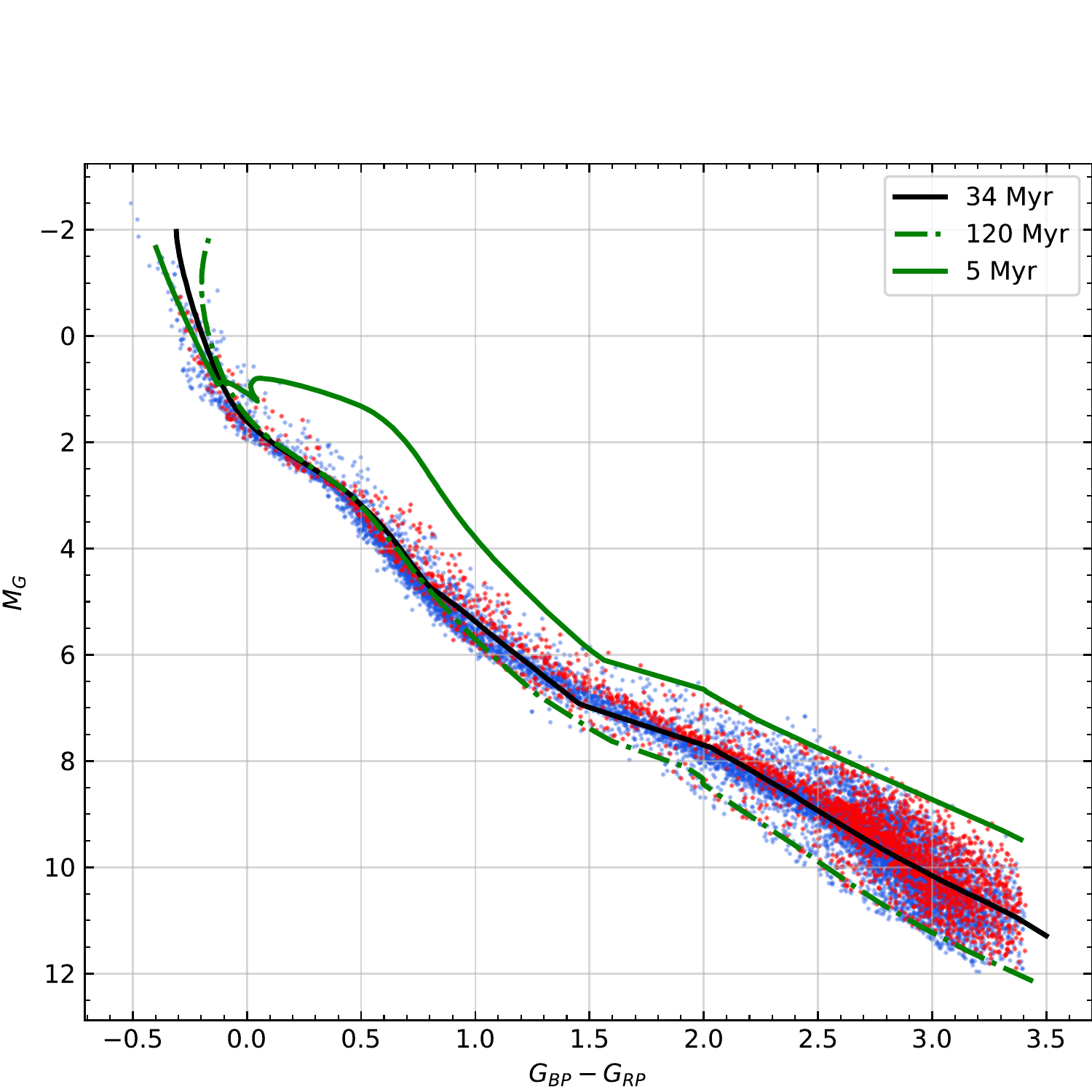}
  \caption{CMDs of the 2694 member candidates in Part\,I (red dots), the 9615 member candidates in Part\,II (blue dots). The black solid curve is the isochrone with an age of 34\,Myr from T20. The other two isochrones are used to remove outliers with ages younger than 5\,Myr (green solid curve) and 120\,Myr (green dot-dashed curve) from the member candidates. All the isochrones are with Z\,=\,0.02. The ages of both Parts\,I and II are 30$-$40\,Myr, although Part\,I seems to be few Myrs younger than Part\,II.
  }
  \label{fig:cmd}
\end{figure}

Figure \ref{fig:cmd} displays the CMDs of the member candidates in Part\,I (red dots) and Part\,II (blue dots) with three isochrones, i.e., 34\,Myr (black solid curve), 5\,Myr (green solid curve), and 120\,Myr (green dashed curve). All three isochrones are from PARSEC \citep{Bressan2012,Chen2014,Chen2015,Tang2014,Marigo2017} CMD\,3.1\footnote{http://stev.oapd.inaf.it/cgi-bin/cmd/3.1} with metallicity $Z=0.02$. As shown in the figure, Parts\,I and II have nearly the same distribution in the CMD, which demonstrates that Parts\,I and II are of similar stellar population and \thj{very likely} one sizable stellar structure which probably originated from the same GMC. We notice that the 34\,Myr isochrone well-matches the observed CMDs for the bright stars, but in the faint end, the isochrone is not perfectly consistent with the observed distribution, particularly for the PMS stars with $M_{G}\gtrsim9$\,mag. This phenomenon was also seen by \citet{Lilu2020ApJ}. This discrepancy is not likely due to the effect of extinction coefficients which should depend on spectral type \citep{Chen(2019)}; accounting for such a discrepancy requires an extinction of $A_{V}>2$\,mag. Most of our sample stars are closer than 400\,pc to the Sun (see Figure \ref{fig:drv}), which is not likely to have such a large extinction value. Given that stars with $M_{G}\gtrsim9$\,mag in our sample have a stellar mass $\lesssim0.5$\,$M_{\odot}$ and $T_{\rm eff}\lesssim3200$\,K, we argue that atmospheric models and PMS stellar evolutionary models of very low mass stars at low temperature need to be further improved. We also find that the sequence of Part\,I is slightly over that of Part\,II. This indicates that the stars in Part II are probably born a few Myr before those in Part\,I if the extinction values are calibrated correctly for both Parts.
%It demonstrates again that Parts\,I and II belong to one huge stellar structure, which probably originates from the same GMC. In addition, the isochrone of 34\,Myr can well match the observed CMDs for the bright stars, but in the faint end, the isochrone is not perfectly consistent with the observed distribution, particularly for the PMS stars. This phenomenon was also found in \citet{Lilu2020ApJ}. Interestingly, the sequence of Part\,I is slightly over that of Part\,II. This indicates that the stars in Part\,II are probably born few Myr before those in Part\,I if the extinction values are calibrated correctly for both Parts.
Slight differences in age are consistent with the differences in distance and proper motion detected in the member candidates of the Snake (Section \ref{sec:motions}).

\begin{figure*}[!htbp]
  \centering
  \includegraphics[width=1.0\textwidth]{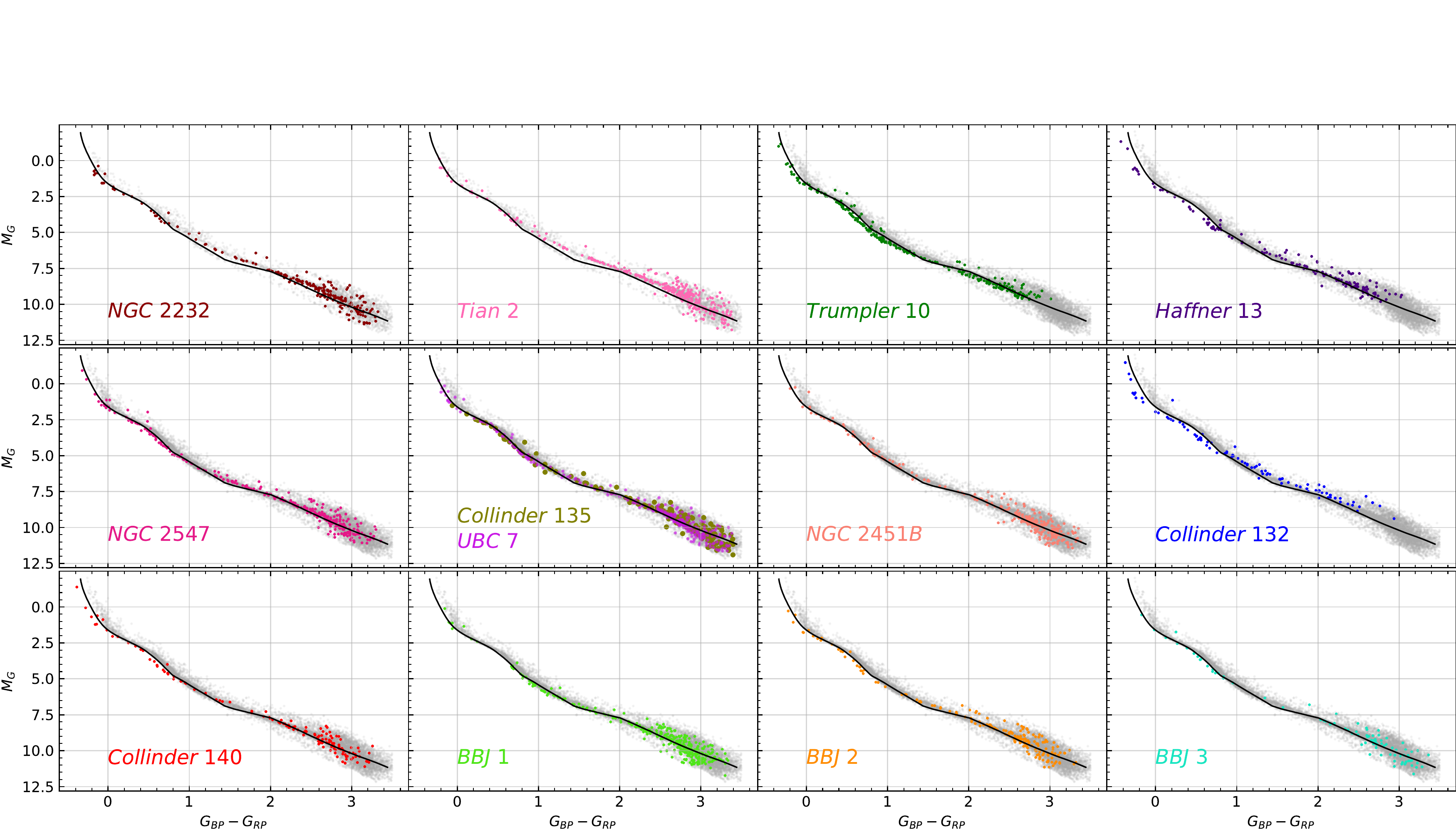}
  \caption{CMDs for the 13 open clusters (same color-coding as in Figures \ref{fig:xyz} and \ref{fig:vlb}) of Parts\,I and II. The gray dots in the upper most row far left two panels and the remaining ten panels are the full sample of 2694 and 9615 member candidates of Parts\,I and II, respectively. The isochrones in all panels are of the same age (34\,Myr) and $Z=0.02$. Trumpler\,10 seems to be older than the other open clusters, which suggests an age of $40-50$\,Myr. The two clusters of the physical pair (i.e., Collinder\,135 and UBC\,7) with exactly same age are displayed in the same panel. According to the CMDs, Tian\,2 and NGC\,2232 of Part\,I are probably also a physical pair.
  }
  \label{fig:cmd_each}
\end{figure*}

Figure \ref{fig:cmd_each} illustrates the CMDs for the 13 open clusters (color-coded). The gray dots in the two left most panels in the upper row are the 2694 member candidates of Part\,I, while the gray dots in the remaining ten panels are the 9615 member candidates of Part\,II. The black curve in each panel is the 34\,Myr isochrone. The CMD of each open cluster well traces the locus of the gray dots, and basically matches the isochrone (black curve), except for the Trumpler\,10. Such discrepancy  indicates that Trumpler\,10 is slightly older than the other open clusters, its age being 40$-$50\,Myr. Compared to previous studies estimating the age of Trumpler\,10, our estimated age is most similar to \citet[][45\,Myr]{Bossini(2019)}, and older than the ages of \citet[][24\,Myr]{Kharchenko(2013)} and \citet[][35\,Myr]{Dias(2002)}. Besides Trumpler\,10, all remaining open clusters have the same age, consistent with the estimated age of the Snake by T20, i.e., 30$-$40\,Myr.

Based on the best fit isochrone, i.e., a 34\,Myr isochrone, we estimate the stellar mass for each member candidate, and find that the new results from both Parts\,I and II support the claims about the mass distribution presented by T20, i.e., more than 84.0\% of the members do not pass the TOn point ($G_\mathrm{BP}-G_\mathrm{RP}<1.0$\,mag) and enter the MS stage, and most of the members have masses with $M<1.0\,\mathrm{M}_{\odot}$. Using the initial mass function of \citet{Kroupa(2001a)}, we roughly obtain a mass of $\sim2500\,\mathrm{M}_{\odot}$ ($\sim8000\,\mathrm{M}_{\odot}$) for Part\,I (Part\,II). Therefore, the total mass of the Snake should be larger than 10,000\,$\mathrm{M}_{\odot}$.

%best fit isochrones (black solid curve and black dashed curve) in the CMD. The result show that the age of Part\,I and Part\,II is around 34\,{\tt Myr}, and the corresponding metallicities are around 0.020. We adopt ages for Part\,I and Part\,II from previous studies when they are in good agreement with locations of the members in the CMD. Such a similar age in 5\,{\tt Myr} and 120\,{\tt Myr} allow us to probe the influence of the secular dynamical evolution of the two parts and interaction with their environments on their morphology (see Section\,5.2). The spectroscopic data favour two larger metallicities, which have been discussed in Section\,4.2. In this panel, we also illustrated two other isochrones with the age of 120\,{\tt Myr} (green dashed) and 5\,{\tt Myr} (green dashed-dot) which are used to remove the obvious outliers from the member candidates. In the bottom-left of panel, there are a few stars which may be some dwarf stars or binary systems.

%----
%\begin{figure*}[!htbp]
%  \centering
%  \includegraphics[width=0.95\textwidth]{fig/mass_20210423_the_0.pdf}
%  \caption{
%  }
%  \label{fig:mass}
%\end{figure*}
%----

%----
%----
%

%----
%----
%\section{}
%\label{}

\subsection{Orbits and Integrals of motion for the open clusters}
\label{sec:orbit}

\begin{figure*}
  \centering
    \includegraphics[width=0.45\textwidth]{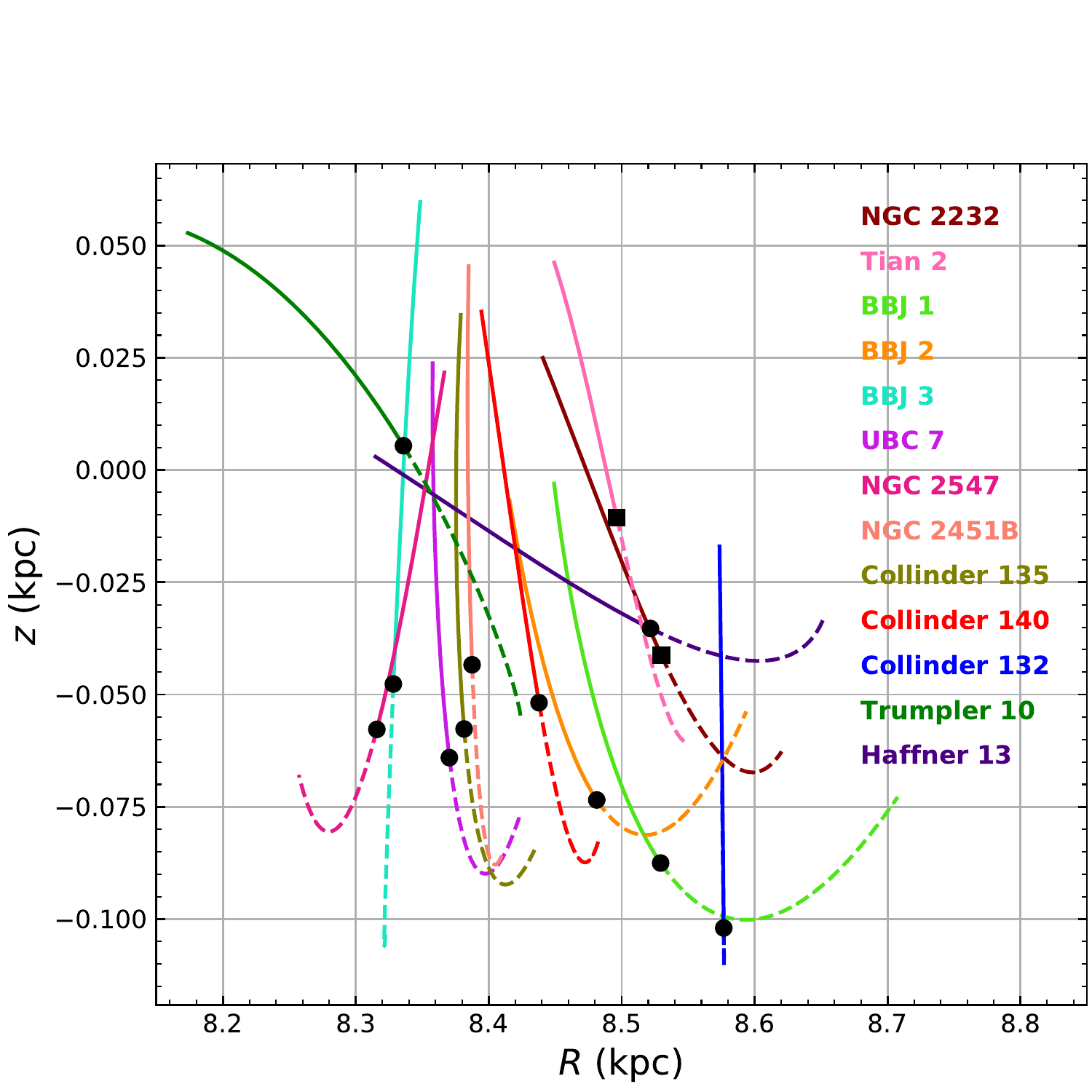}
    \includegraphics[width=0.45\textwidth]{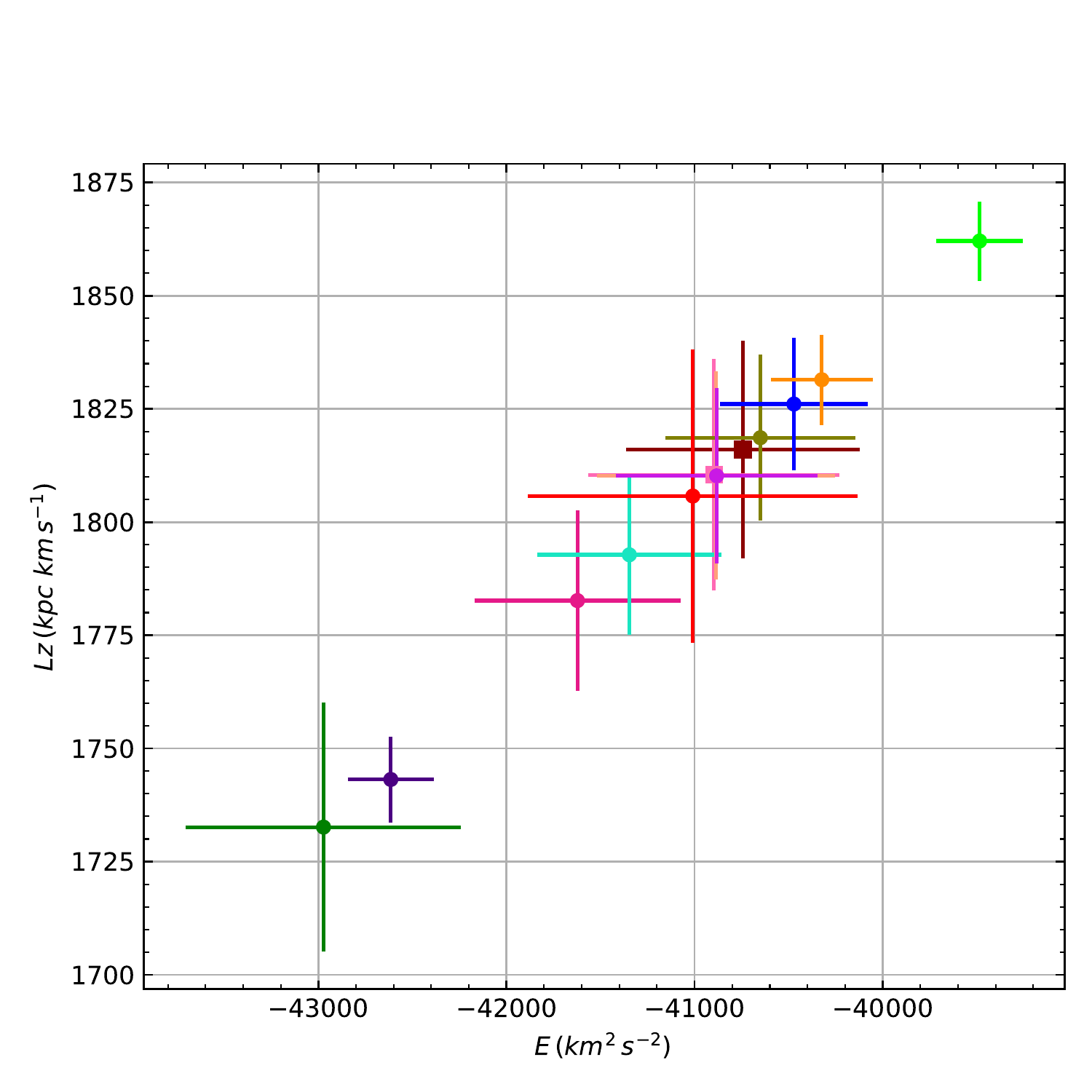}
  \caption{Left panel: Orbits of the 13 open clusters. The orbital parameters of each open cluster are the median values of its member candidates with $R_V$. Each orbit includes three parts:current positions \wf{(Part\,I: black filled squares, Part\,II: black dots)}, past orbits 40\,Myr ago (solid curve), and future orbits 50\,Myr later (dashed curve). \wf{Right panel: Distribution of the 13 open clusters in the integrals of motion ($L_Z$ and $E$) space. The filled squares (Part\,I) or dots (Part\,II) with error bars illustrate the median and {\tt rms} values of $L_Z$ and $E$ of the member candidates with $R_v$ in the clusters.} The color-coding is the same as in Figures \ref{fig:xyz}, \ref{fig:vlb}, and \ref{fig:cmd_each}.
  }
  \label{fig:orbit and angular momentum}
\end{figure*}

Most of open clusters include at least 10 member candidates with $R_V$, except for Haffner\,13, Collinder\,132 and BBJ\,1 which only have 5, 6 and 8 members with $R_V$, respectively. We therefore have a sufficient number of open cluster member stars to calculate the central position ($l_0$, $b_0$) and the median value of $R_V$, proper motion, distance from the Sun, and $Z$-component of angular momentum ($L_Z$) for each cluster. \wf{We use the potential model of \citet{Bovy(2015)} and the 6D phase information to calculate the orbits and total energy $E$ for the 13 open clusters. Figure \ref{fig:orbit and angular momentum} shows the color-coded orbits in the $R-Z$ space (left panel) and integrals of motion $E-L_Z$ (right panel) for the 13 open clusters.}%(middle)} and $L_Z-L$ (right) panels.} 

In the left panel, each orbit includes three parts: (1) a black dot or filled square that is the current location of the cluster; (2) the solid curve which is the past orbit over the last 40\,Myr; and (3) the dashed curve which is the future orbit over the next 50\,Myr. It \wf{looks} that in the past most of the open clusters were physically closer together compared to their present locations, and move away from each other in the future. This indicates that the whole structure is expending. Trumpler\,10 and Haffner\,13 present slightly different orbits compared to the other clusters.

In the right panel, \wf{Tian\,2 and NGC\,2232 in Part\,I and most components of Part\,II are assembled together and almost indistinguishable. This suggests that Parts\,I and II have considerably similar integrals of motion which demonstrates that Part\,I and most components of Part\,II are very likely one single unified structure. It worth noting that, Trumpler\,10 and Haffner\,13 which locate in the lower left part of the figure, are similar to each other but with different integrals of motion from most of the other clusters of Part\,II. Interestingly, BBJ\,1 is also slightly isolated from the majority of clusters in Part\,II. There are two possible reasons: i) the integrals of motion of BBJ\,1 are calculated only from 8 member candidates with radial velocities, the sample is sparse thus harm the significant level of its statistics. ii) some local regions including BBJ\,1, Trumpler\,10 and Haffner\,13 probably experienced some unknown perturbations after they were formed.} 

%, the Z component angular momentum and total energy distribution of the clusters are nearer and further we can infer that they are more similar in dynamics \citep{Helmi2000}. 
%It is the reason why the three clusters Trumpler\,10, Haffner\,13 and BBJ\,1 slightly differ with other clusters in this figure and we see there are some differences with 3D velocity in Table.} 

%----

\subsection{3D expansion rates}
\label{sec:3d_ex_r}

The orbits (Section \ref{sec:orbit}) of the 13 open clusters demonstrate that the whole structure is expanding. In this section, we illustrate how
localized regions (e.g., open clusters) expand. In Section \ref{sec:open_clusters}, we divided the whole structure into 5 groups, and obtained the 6D phase information for some group members, i.e., 196, 159, 369, 264, and 7 stars with radial velocities in Groups\,I$-$V, respectively. In the following, we will use these stars to explore the 3D kinematics of Group\,I-IV. Because Group\,V only includes 7 stars with radial velocities, it is difficult to statistically study. We therefore do not calculate the expansion rate for Group\,V. We divide the member candidates of each group into 5 bins in the $X$-, $Y$-, and $Z$-directions, respectively. We consider the stars which are beyond $\pm3\sigma$ of the corresponding velocities as outliers in each bin, and remove the outliers before the statistical analysis.

%and positions ($l$,$b$) and then validate the kinematic structure of the five groups by establishing whether they exhibit a spatial velocity gradient with the whole structure included Part\,I and Part\,II. And group\,I hosts Trumpler 13, and group\,II is made up of BBJ\,1 and BBJ\,2, and group III contains NGC 2547, NGC2451B, BBJ3, Collinder 135, UBC 7, Collinder 140, and group\,IV consists of NGC 2232 and Tian 2. We can see the distinctions of them in the above figures 1 to 3. A positive correlation between the position along a given axis and velocity along this axis is interpreted as evidence for expansion. The approach we use is similar to that followed by \citet{Wright(2018)} to the Scorpius-Centaurus OB association. 

%The most direct way to study the internal kinematics of a group of stars is to independently determine a 3D velocity vector for each member star. We use the radial velocities of each group from {\tt Gaia EDR3} with a typical uncertainty of 3.3\,$km s^{-1}$. We calculated the position (X, Y, Z) in Galactic Cartesian coordinates for each member of each group, and its velocity vector (U, V, W) in Section\,4.1. For each group, we discarded the member candidates whose velocity was not the sample mean by more than $3\sigma$, and then fit a linear relationship between X and U (and similarly for the Y and Z axes). The slopes $(\kappa_X, \kappa_Y, \kappa_Z)$ we derived are listed in Table\,1 and are shown in Figure\,14. 

%----
\begin{figure*}[!htbp]
  \centering
  \includegraphics[width=0.95\textwidth, trim=0.1cm 0.7cm 0.0cm 0.8cm, clip]{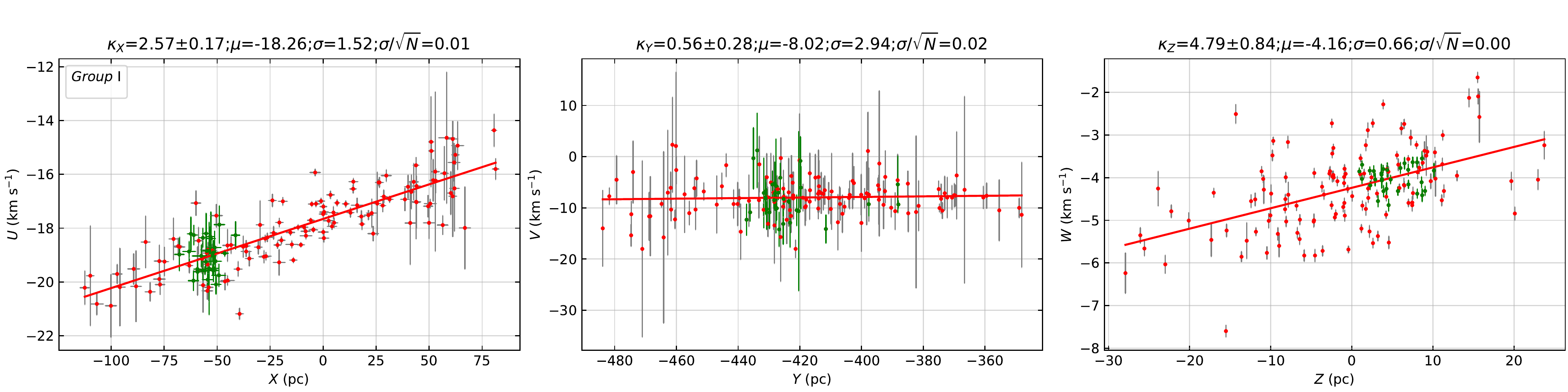}
  \includegraphics[width=0.95\textwidth, trim=0.1cm 0.7cm 0.0cm 0.3cm, clip]{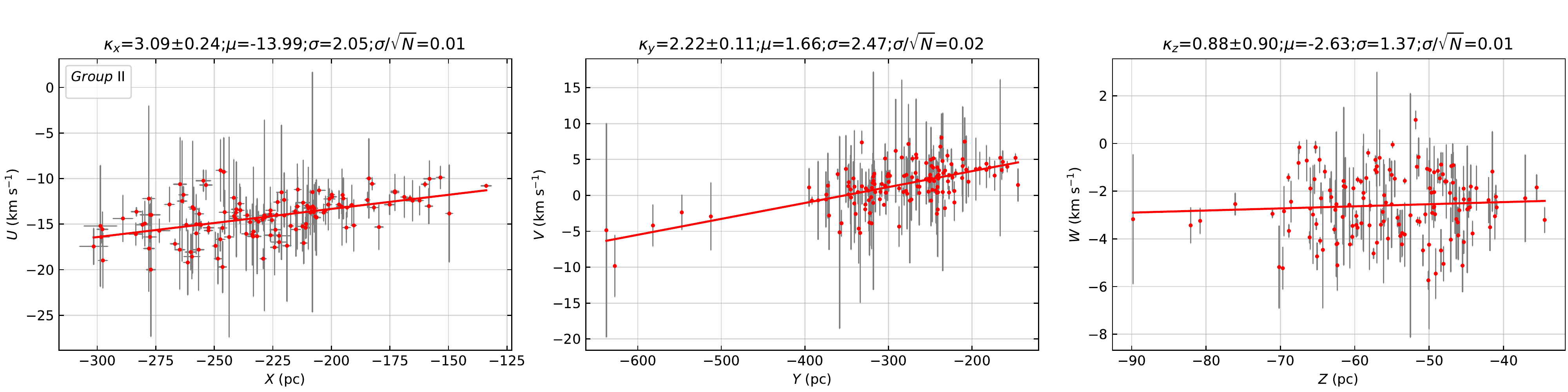}
  \includegraphics[width=0.95\textwidth, trim=0.1cm 0.7cm 0.0cm 0.3cm, clip]{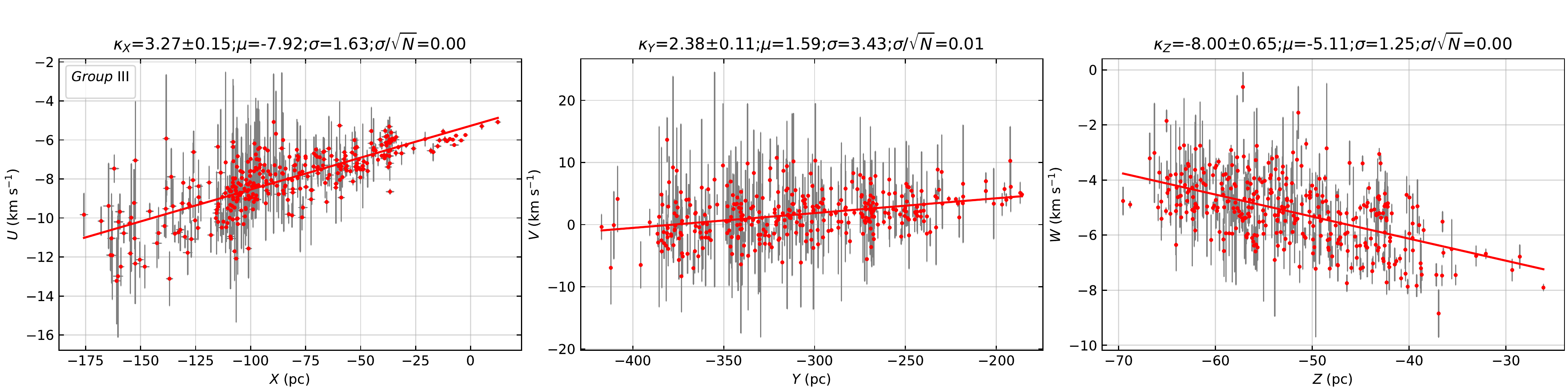}
  \includegraphics[width=0.95\textwidth, trim=0.1cm 0.1cm 0.0cm 0.3cm, clip]{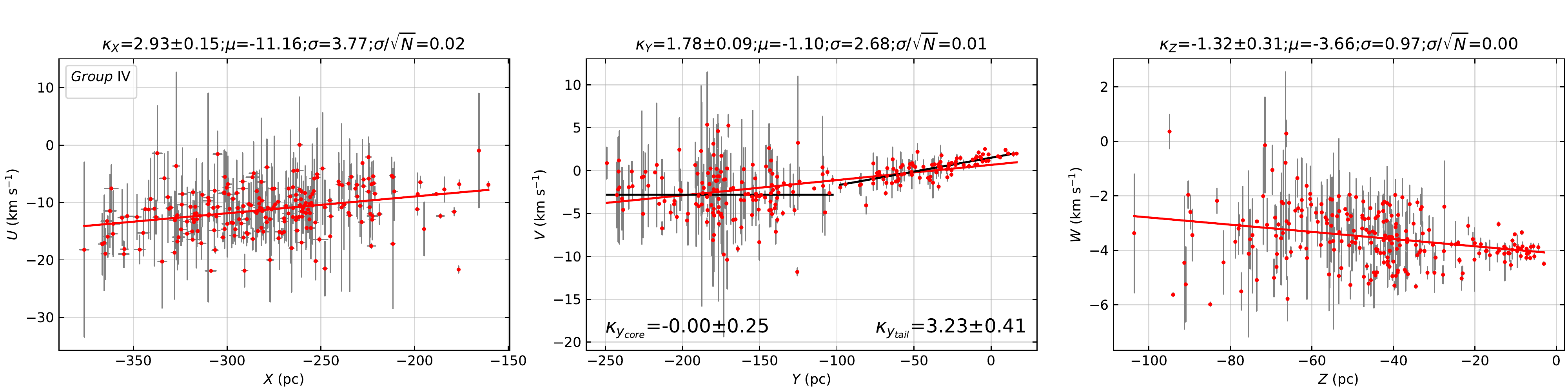}
  \caption{3D velocities ($U,V,W$) as functions of 3D positions ($X,Y,Z$) for stars (red dots) in Group\,I$-$IV. The solid lines show the best fit linear relationships between the plotted quantities. The best fitting slope $\kappa$ with its uncertainty ($\times10^{-2}$\kms$\rm pc^{-1}$) and mean velocity $\mu$ (\kms) with its standard deviation $\sigma$ (\kms) and statistical uncertainty $\sigma/\sqrt N$ (\kms) are noted above each panel. The green dots are the member candidates of Trumpler\,10. If the groups are not expanding we expect slopes of zero, while if they are expanding we expect a positive slope (a negative slope implies a sign of contraction along the corresponding axis). Interestingly, Group\,IV presents two clear modes in the relationship between $Y$ and $V$ (lowest row, middle panel). The two clusters in the core (i.e., Tian\,2 and NGC\,2232 of Part\,I) show neither sign of expansion nor contraction ($\kappa_{Y_\mathrm{core}} \simeq0.0\times10^{-2}$\kms$\rm pc^{-1}$) at $Y<-100$\,pc; while the tail at $Y>-100$\,pc clearly shows signs of expansion ($\kappa_{Y_\mathrm{tail}} \simeq3.2\times10^{-2}$\kms$\rm pc^{-1}$), as shown by the two black regression lines (lowest row, middle column panel). Group\,III presents a significant sign of contraction in the $Z$-direction ($\kappa_Z \simeq-8.0\times10^{-2}$\kms$\rm pc^{-1}$). All slopes are summarized in Table \ref{tab:sums}.}
  \label{fig:expansion}
\end{figure*}
%----

Figure \ref{fig:expansion} displays the 3D kinematical variations along the 3D configuration space of the member candidates of Groups\,I$-$IV (upper to lower rows, respectively). The panels show $X-U$ (first column), $Y-V$ (second column), and $Z-W$ (third column). All four groups present significantly linear correlation between $X$ and $U$ with almost the same slope (i.e., $\kappa_X \simeq3.0\times10^{-2}$\kms$\rm pc^{-1}$). This indicates that all of the four groups are expanding in the $X$-direction. According to the empirical relationship between the slope $\kappa$ and the expansion age, i.e., $\tau=(\gamma\kappa)^{-1}$ (where $\gamma$\,=\,1.0227 is the conversion factor from \kms\ to pc\,Myr$^{-1}$ \citep{Blaauw1964,Wright(2018)}), $\kappa_X \simeq3.0\times10^{-2}$\kms$\rm pc^{-1}$ signifies that the expansion age is around 33\,Myr, which is highly consistent with the age of the Snake ($34$\,Myr). 

In the phase space $Y-V$, Group\,II and III present significant expansion similar to the phase space $X-U$; the values of $\kappa_Y$ are $2.22\pm0.11$ (for Group\,II) and $2.38\pm0.11$ (for Group\,III) $\times10^{-2}$\kms$\rm pc^{-1}$. Group\,I does not show a sign of expansion in the $Y$-direction ($\kappa_Y \simeq0.56\pm0.28\times10^{-2}$\kms$\rm pc^{-1}$). Interestingly, Collinder\,132 is far from BBJ\,1 and BBJ\,2 in Group\,II, but its expansion rate (the five points at $Y<-500$\,pc) seems to be well matched with the other Group\,II members. Group\,IV displays two clear modes in the relationship between $Y$ and $V$: the two clusters (i.e., Tian\,2 and NGC\,2232) show no sign of expanding or contraction (i.e., $\kappa_Y \simeq0.0\pm0.25\times10^{-2}$\kms$\rm pc^{-1}$) at $\rm Y<-100$\,pc; while the tail ($\rm Y>-100$\,pc) clearly shows signs of expanding ($\kappa_Y \simeq3.2\pm0.41\times10^{-2}$\kms$\rm pc^{-1}$), as shown by the two black lines (fourth row from the top, middle panel).

In the phase space $Z-W$, only Group\,I displays a sign of expansion ($\kappa_Z \simeq4.79\pm0.84\times10^{-2}$\kms$\rm pc^{-1}$). Strangely, Group\,III exhibits a strong sign of contraction in the $Z$-direction ($\kappa_Z \simeq-8.0\pm0.65\times10^{-2}$\kms$\rm pc^{-1}$). This phenomenon was also noted by CG19b who measured a slope of $\kappa_Z \simeq-4.49\pm0.37\times10^{-2}$\kms$\rm pc^{-1}$. Group\,II and IV do not display any significant signs of expansion or contraction. 

Taking Trumpler\,10 in Group\,I as an example, we investigate how the elongation effect along the line of sight as induced by inverting parallax (such that distances are $1000\omega^{-1}$ (pc)) affects the calculated expansion rate of the cluster. As shown by the green dots in Figure \ref{fig:expansion}, the elongation effect mainly increases the uncertainty of velocity components in a limited region, but does not significantly affect the expansion rate.

The slopes for the four groups are summarized in Table \ref{tab:sums}.

%And we see that all four groups give a import positive correlation between X and U ($\kappa_X$\,>\,0). Most of them also give signs of expansion along Y axis, with the exception of group\,I, whose velocity gradient $\kappa_Y$ appears negative. As for Z axis (perpendicular to the Galactic plane), some of them exhibit positive, with the exception of group\,II and III, which give signs of contraction. Following the formulation of Blaauw(1964) and \citet{Wright(2018)}, the slope $\kapppa$ can be also expressed as an expansion age: $\tau=(\gamma\kappa)^{-1}$ (where $\gamma$\,=\,1.0227 is the conversion factor from \kms\ to {\tt pc/Myr}. The 0.03

%Inverting this relation allows us to derive $\kappa$ from an estimated age.

\subsection{Gas Environment}
\label{sec:gas}

\begin{figure*}[!htbp]
  \centering
 \includegraphics[width=1\textwidth]{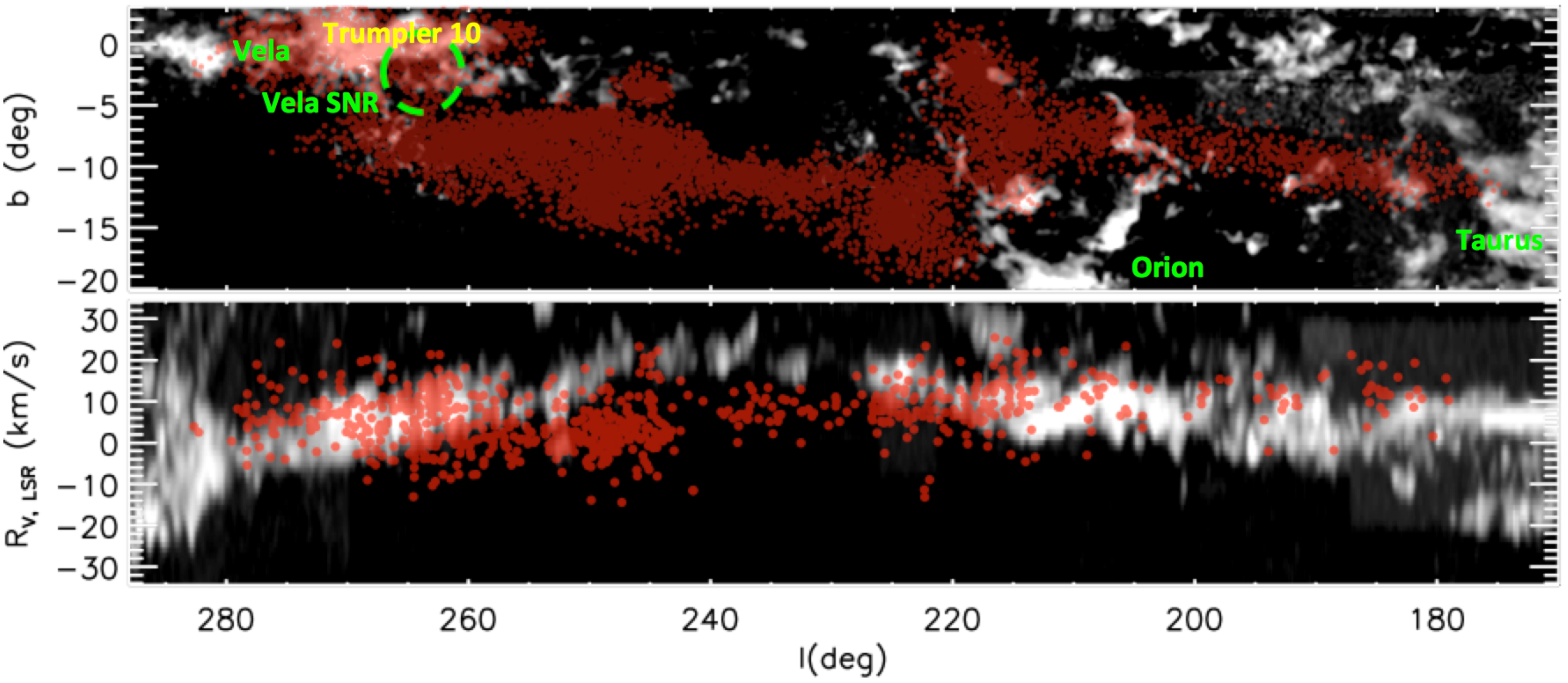}
  \caption{Distribution of the member candidates (red filled circles) of the Snake and {\tt CO} gas (white) integrated over the velocity range between $-$14 and 25\,\kms\ chosen according to the $R_{V,\mathrm{LSR}}$ range of the stars in the panels of $l-b$ (top panel) and $l-R_{V,\mathrm{LSR}}$ (bottom panel). The radial velocities of both the member candidates and gas are with respect to the LSR. The green dashed circle displays the location of the Vela supernova remnant (Vela SNR).
  }
  \label{fig:gas}
\end{figure*}

%Young stars are usually grouped in clusters and located within their natal star-forming regions, and in contrast to younger stars, older stars are found dispersed throughout the Galactic filed \citep{Swiggum(2021)}.
To investigate the relationship between the Snake and its surrounding gas environment, Figure \ref{fig:gas} illustrates the overlapping distributions of the member candidates of the Snake and {\tt $^{12}$CO} integrated intensity map in the velocity range from between $-$14 and 25\,\kms\ in $l-b$ (top panel) and $l-R_{V,\mathrm{LSR}}$ (bottom panel). The {\tt $^{12}$CO} data are taken from \cite{Dame2001}, and the velocity range is chosen in accordance with the radial velocity range of the member candidates. Radial velocities of both the member candidates and the gas are with respect to the LRS. It seems that most of the components are not associated with the {\tt $^{12}$CO} molecular cloud, which is expected since the gas has already likely dispersed in the past 30$-$40\,Myr.

 In Figure \ref{fig:gas}, the candidate members in Trumpler\,10 seem to show a spatial and kinematic correlation with the {\tt $^{12}$CO} molecular gas in their localized region. However, given that Trumpler\,10 has an age of $\sim$40$-$50\,Myr and the molecular cloud may not have remained present for so long \citep{Hartmann2001},   Trumpler\,10 is very unlikely associated with the  {\tt $^{12}$CO} molecular cloud. Actually, the {\tt $^{12}$CO} molecular cloud shown in the region of Trumpler\,10 is the Vela molecular cloud \citep{1991A&A...247..202M} with a distance of $\sim870-970$\,pc \citep{Zucker2020}. Thus, there is no physical association between Trumpler\,10 and the Vela molecular cloud.

 The Vela supernova remnant (Vela SNR), one of the closest SNRs to us (287$^{+19}_{-17}$\,pc, \citealt{Dodson2003ApJ}), is located in the sky region nearby Trumpler\,10, as shown by the dashed circle in Figure \ref{fig:gas}. The Vela SNR interacts with its surrounding structures, e.g., the IRAS Vela Shell (IVS) which a ring-like structure discovered in far-infrared images by \citet{Sahu(1992)} and the massive binary system $\gamma^{2}$\,Velorum ($336^{+8}_{-7}$\,pc, \citealt{North2007MNRAS}). As modeled by \citet{Sushch2011}, the IVS is the result of the stellar-wind bubble of $\gamma^{2}$\,Velorum, and the observed asymmetry of the Vela SNR is due to the envelope of the Vela SNR physically meeting with the IVS. Furthermore, CG19a proposed that the stellar feedback and supernovae which exploded in the 30\,Myr old clusters swept the gas away and produced the IVS, and that the expansion of the IVS triggered a second burst of star formation (e.g., $\gamma^{2}$\,Velorum) approximately 10\,Myr ago, as shown in their Figure 9. Our results suggest that Trumpler\,10 seems unlikely to have participated in such a joint evolutionary process, since its distance is around 430\,pc (see Table \ref{tab:sums}), which is much larger than that of the Vela SNR. However, the nearby clusters (e.g., NGC\,2547, BBJ\,3, and NGC\,2451B) have very similar distances with the Vela SNR. They are probably associated with the Vela SNR.

%\begin{figure*}[!htbp]
%  \centering
% \includegraphics[width=1\textwidth]{fig/lxlylz_20211029_the_0.pdf}
%  \caption{Distribution of the member candidates (red filled circles) of the Snake and {\tt CO} gas (white) integrated over the velocity range between $-$14 and 25\,\kms\ chosen according to the $R_{V,LSR}$ range of the stars in the panels of $l$-$b$ (top) and $l$-$R_{V,LSR}$ (bottom). The radial velocities of both the member candidates and gas are with respect to the LSR. The green dashed circle displays the location of the Vela supernova remnant (Vela SNR).
%  }
%  \label{fig:angular mementum}
%\end{figure*}

\section{Discussion}
\label{sec:discu}
In this section, we detail the likely formation scenario for the whole system, the Snake and its surrounding territory, and then briefly compare with other works in the literature. 

%except the two open clusters, i.e., Trumpler\,10 and Haffner\,13 

%Associations especially those nearby associations are the ideal laboratory for studying stellar kinematics and evolution of stars in groups, therefore, efforts have been taken for searching more new associations, especially after the data release of Gaia \citep{Liu(2020)}.

%They are probably formed from a giant molecular cloud, but

\subsection{Origin of the structures}
\label{sec:origin}
We have investigated the properties of the member candidates for Parts\,I and II. In the perspective of spatial distributions, Parts\,I and II are well bridged, except for Haffner\,13 \wf{and Collinder\,132}, which are relatively further than the other components. %parts.
%\wf{We calculated the average proper motions of Haffner\,13 and Collinder\,132: 17.39$\pm$1.85\,\kms\ and 12.29$\pm$0.76\,\kms\ separately, and the difference between the two open clusters (Haffner\,13 and Collinder\,132) and others (excluded Haffner\,13 and Collinder\,132) is a few\,\kms\. The mean distances of them are 558.7$\pm$20.8\,pc and 649.1$\pm$41.6\,pc, separately. And the average distance of Part\,I is 319.2$\pm$42.8\,pc, and Part\,II’s (excluded Haffner\,13 and Collinder\,132) is 370.6$\pm$57.7\,pc. The difference is about 51\,pc. The age of the 13 open clusters, included Haffner\,13 and Collinder\,132,is the range of $30-40$\,Myr and similar to others'. As shown as Figure 11, we go back 40\,Myr and see that all clusters have a similar spatial position. Although we can see Figure \ref{fig:angular mementum} that Haffner\,13 and Collinder\,132 have bigger differences than others, all clusters have a similar angular momentum distribution. Besides, according to \citet{Beccari(2020)}, we can get that this is an uncovering a 260\,pc, 35\,Myr old filamentary relic, which included Haffner\,13 and Collinder\,132 in Figure 6. From above, Haffner\,13 and Collinder\,132 have a possible to be parts of Parts\,I and\,II. We think they are interesting and hope that we can get more information, for example, more radial velocities and metallicities, in the future and will further study and discuss whether they might have a relation with others or not.}
\wf{Parts}\,I and II have nearly the same ages (i.e., $30-40$\,Myr), even though Part\,II appears a few Myr older than Part\,I in the CMDs. Meanwhile, Trumpler\,10 appears to be around 10\,Myr older than the other parts. \wf{Considering integrals of motion, Part\,I has considerably similar $L_z$ and $E$ with most components of Part\,II.} We do not detect significant differences in the metallicities of the two parts. With these in mind, \wf{we propose that Parts\,I and II are very likely} born from the same environment, and belong to one sizable structure, but exhibit different formation \wf{and evolution} history in some localized regions.

\thj{Though Haffner\,13 and Collinder\,132 are not well bridged with other components in Part\,II, and Haffner\,13, Trumpler\,10, and BBJ\,1 present slightly different integrals of motion from the remaining components of Part\,II, we argue that the four clusters most likely belong to the whole structure of Part\,II, as claimed by BBJ20, or at lease were born from very similar environments in a very similar epoch, because the clusters share similar ages and metalicities with the other components of Part\,II.}

In T20, we have basically ruled out the formation channel of tidal tails for such a sizable structure. The population is so young (only 30$-$40\,Myr) that it cannot be well explained with the classical theory of tidal tails \citep{Kharchenko2009,Jerabkova2019,Jerabkova2021}. The reasonable explanation is that the Snake is a hierarchically primordial structure, probably formed from a filamentary GMC. Gravitationally bound stellar clusters in this structure arise naturally at high local density regions, while unbound associations form in situ at low density regions. Therefore, the most probable scenario is like this: the older open cluster (e.g., Trumpler\,10) was firstly born in the GMC around 40$-$50\,Myr ago, and then the filamentary sub-structures in Part\,II including the 10 open clusters (e.g., a physical pair of Collinder\,135 and UBC\,7, etc.) were formed around 10\,Myr later. After a few Myrs, the sub-structures in Part\,I including another physical pair of open clusters (i.e., Tian\,2 and NGC\,2232) were formed. During the Snake's evolution, stellar feedback and supernovae of massive OB stars that exploded in the old (30$-$40\,Myr) clusters may have produced a central cavity and shell, e.g., IVS \citep{Sahu(1992)}, and triggered a second burst of star formation \citep[e.g., the 3$-$4\,Myr binary system $\gamma^2$\,Velorum,][]{Jeffries(2009)} $\sim$10\,Myr ago (CG19a). \thj{In addition, \citet{Tutukov2020} proposed that the snake-like or stream-like stellar structures possibly formed due to the decay of star clusters and
OB associations.}

The whole structure is expanding, as well as localized regions such as the open clusters, particularly in the $X$-direction. The expansion should not be due to the stellar winds from the massive OB stars, but caused by an event that took place before the stellar members formed, because the entire Snake is expanding, as discussed in CG19a.

\subsection{Comparison with other works}
\label{}

BBJ20 applied the algorithm {\tt DBSCAN} on {\tt Gaia DR2}, and found a 260\,pc wide and 35\,Myr old filamentary structure nearby the Snake of T20. To study the relationship between the Snake and BBJ20's structure, we build two samples (i.e., Parts\,I and II) to totally cover these two structures and run the coherent FOF algorithm as employed by T20 on the most recent data release of {\tt Gaia} ({\tt Gaia EDR3}). Using FOF we are able to recover the two structures and find that they are well bridged. Comparing with BBJ20, we find more fine sub-structures. For example, a physical pair of open clusters \citep{Kovaleva(2020c)}, Collinder\,135 and UBC\,7, can be clearly distinguished in our newly combined structure (see Figure \ref{fig:lb_pm}).

CG19b investigated the 3D expansion rates of the young population in the Vela-Puppis region. In their work, the authors divided the young population into seven sub-populations. The members of both their Population\,II and our Group\,I mostly belong to Trumpler\,10. Their Population\,IV member mainly belong to Collinder\,135, UBC\,7, Collinder\,140, NGC\,2547, and NGC\,2451B, while our Group\,III includes these plus an addition open cluster BBJ\,3. Because the algorithms and the data (CG19b used {\tt Gaia DR2}) are different, the final member candidates are slightly different. Taking Trumpler\,10 as an example, CG19b find several sparse member candidates at $l<255$\degr (see their Figure 3), but we find no such members in this region. We cut the extra member candidates from the sample of CG19b, and carefully compare them with our sample. However, we find the extra candidates do not match well with our sample's kinematics or CMD. Despite such differences in samples and sample selection between CG19b and our work, we find that the expansion rates are roughly consistent with each other (we compare our results in Table \ref{tab:sums}).

\section{Summary and Conclusions}
\label{sec:discu}
To search for all the stellar members belonging to the Snake and reveal its complete structure, we build two samples (i.e., Parts\,I and II) using {\tt Gaia EDR3} to investigate the relation between the Snake as defined by T20 and the filementary structure found by BBJ20. With the coherent FOF algorithm, we identified 2694 and 9615 member candidates from Parts\,I and II, respectively. Embedded within these member candidates are 13 open clusters. By combining spectroscopic data from multiple surveys, we investigate the comprehensive properties of these member candidates and find that they \thj{are very likely to} belong to one sizable structure, \thj{because (1) most of the components are well bridged in configuration space (except for two distant clusters Collinder\,132 and Haffner\,13), (2) most of the components follow a single stellar population with an age of 30$-$40\,Myr (\thj{except for the slightly older cluster Trumpler\,10}) and share solar metallicity, and (3) Part\,I presents considerably similar orbits and integrals of motion with most components of Part\,II.}

%The components (e.g., the 13 open clusters) present different kinematics.
The orbits of the 13 open clusters indicate that the whole structure is expanding. According to the tangential velocities $v_{l}$ and $v_b$, we divide our sample into five groups. We then calculate the 3D expansion rates for four of the five groups (Group\,V includes too few member candidates with $R_V$ to accurately measure its expansion rate). The expansion rates show that the four groups are expanding at a coherent rate ($\kappa_X\sim3.0\,\rm\times 10^{-2}$ \kms\ $\rm pc^{-1}$) in the $X$-direction. The corresponding expansion age is perfectly consistent with the age of the Snake. Interestingly, Group\,III presents a significant sign of contraction in the $Z$-direction. The mechanism causing the expansion and contraction are still not clear.
%It suggests that the two open clusters indeed are sub-structures, as presented in their spatial distribution and ages

In this work we rule out the formation channel of the Snake as due to a tidal tail. Our work shows that the Snake is most likely a hierarchically primordial structure, and probably formed from a filamentary GMC with different formation histories in several different localized regions. We propose the most likely formation scenario of the Snake is as follows. Trumpler\,10 was first to be born in the GMC around 40$-$50\,Myr ago, followed by the filamentary sub-structures in Part\,II including 10 open clusters (e.g., Collinder\,135, UBC\,7, etc), which formed around 10\,Myr later. After another few Myrs, the sub-structures in Part\,I including Tian\,2 and NGC\,2232 formed.

\thj{Trumpler\,10, Haffner\,13, and BBJ\,1 have slightly different integrals of motion ($L_z$ and $E$) from the other components of the Snake. This indicates that these localized regions probably experienced some type of dynamical perturbation after they were formed.} During the Snake's evolution, stellar feedback and the explosion of supernovae of massive OB stars in the old (30$-$50\,Myr) clusters may have produced a central cavity and shell \citep[e.g., IVS,][]{Sahu(1992)}, and trigger a second burst (e.g., the binary system $\gamma^2$\,Velorum) of star formation $\sim$10\,Myr ago (CG19a).

The Snake has more than ten thousand member candidates following a single stellar population (30$-$40\,Myr old) and continuous mass function. Besides the massive O and B members, it also includes a number of lower-mass stars and PMS stars over hundreds of parsecs in our solar neighborhood. Therefore, it will provide us an ideal laboratory to study the history of stellar formation, environmental evolution, and stellar physics. For example, it is a good sample to study stellar binarity \citep{El-Badry2019MNRAS,Tianetal2020ApJS} in young populations, the surface rotation dichotomy \citep{SunWJ2019ApJ}, the initial mass function of low-mass stars \citep{Naama2021MNRAS}, and so on. Also, it has great potential to reveal the relationship between young stellar populations and the spiral arms \citep{Quillen(2020),HaoCJ2021}, the rotating bar \citep{Tian2017RAA}, and the vertical phase mixing \citep{Antoja(2018),Tian2018ApJ,LiZY2021ApJ}. Moreover, it may provide some clues to understand the formation of some Galactic-scale structures, e.g., the Gould Belt \citep{Poppel1997} and the Radcliffe Wave \citep{Alves2020Natur}.

%==================================================================

%==============================================================
\acknowledgments
We thank Feng Wang, Zhiyuan Ren, Chengyuan Li, Chao Liu, Jing Zhong for helpful discussions and acknowledge the National Natural Science Foundation of China (NSFC) under grant Nos. 11873034, 12011530421, the Cultivation Project for {\tt LAMOST} Scientific Payoff and Research Achievement of CAMS-CAS, and the science research grants from the China Manned Space Project with NO.CMS-CSST-2021-A08. This work has made use of data from the European Space Agency (ESA) mission {\tt Gaia} (https://www.cosmos.esa.int/gaia), processed by the {\tt Gaia} Data Processing and Analysis Consortium (DPAC, https://www.cosmos.esa.int/web/gaia/dpac/consortium). Funding for the DPAC has been provided by national institutions, in particular the institutions participating in the {\tt Gaia} Multilateral Agreement. Guoshoujing Telescope ({\tt LAMOST}) is a National Major Scientific Project built by the Chinese Academy of Sciences. Funding for the project has been provided by the National Development and Reform Commission. {\tt LAMOST} is operated and managed by the National Astronomical Observatories, Chinese Academy of Sciences.
%==============================================================
\software{{\tt Astropy} \citep{AstropyCollaboration2013}, {\tt ROCKSTAR} \citep{Behroozi(2013)}, {\tt Numpy} \citep{vanderWalt2011CSE}, {\tt Matplotlib} \citep{Hunter2007CSE}, {\tt TOPCAT} \citep{Taylor2005ASPC}}.

{\bf Data availability:} The data supporting this article will be shared upon reasonable request to the corresponding author.
\bibliographystyle{aasjournal}
%\bibliography{main}
\bibliography{bibfile}

%--
%{\color{blue}
%\begin{landscape}
\clearpage
\begin{turnpage}

\begin{table*}[htbp] \scriptsize %\tiny
\caption{The properties of the 13 open clusters included in Parts\,I and II.}
\begin{tabular}{c|c|c|c|c|c|c|c|c|c|c|c|c|c|c|c}
\hline
\hline
Samp.& Cluster& $l_0$ & $b_0$ & $d$ & $U$ & $V$ & $W$ & ${\mu_{l^*}}^a$ & ${\mu_b}^a$ & $R_V$& Age & Gr. & $\kappa_{\rm X}$ & $\kappa_{\rm Y}$ & $\kappa_{\rm Z}$\\
\hline
&&\multicolumn{2}{c|}{($\degr$)}&pc&\multicolumn{3}{c|}{\kms}&\multicolumn{2}{c|}{\masyr}&\kms&Myr&&\multicolumn{3}{c}{$\rm 10^{-2}kms^{-1}pc^{-1}$}\\
\hline   %   $\pm$ uncertainty

\multicolumn{1}{c|}{\multirow{2}{*}{Part\,I}}&Tian\,2&218.33&-2.12&$285.9\pm0.3$&-8.0$\pm$0.9&-1.9$\pm$0.7&-4.0$\pm$0.1&-1.1$\pm$0.1&-7.6$\pm$0.1&$21.8\pm1.1$&34&\multicolumn{1}{c|}{\multirow{2}{*}{IV}}&\multicolumn{1}{c|}{\multirow{2}{*}{2.93$\pm$0.15}}&\multicolumn{1}{c|}{\multirow{2}{*}{1.78$\pm$0.09}} &\multicolumn{1}{c}{\multirow{2}{*}{-1.32$\pm$0.31}}\\
\cline{2-12}
\multicolumn{1}{c|}{}&
NGC\,2232&214.42&-7.51&$319.2\pm0.5$&-9.9$\pm$1.1&-2.6$\pm$0.9&-4.0$\pm$0.3&-0.6$\pm$0.1&-5.0$\pm$0.1&$24.7\pm1.4$&34&\multicolumn{1}{c|}{}&\multicolumn{1}{c|}{}&\multicolumn{1}{c|}{}&\multicolumn{1}{c}{}\\
\hline

\multicolumn{1}{c|}{\multirow{12}{*}{Part\,II}}&\multicolumn{1}{c|}{\multirow{2}{*}{Trum.\,10}}&\multicolumn{1}{c|}{\multirow{2}{*}{262.71}}&\multicolumn{1}{c|}{\multirow{2}{*}{0.72}}&\multicolumn{1}{c|}{\multirow{2}{*}{$433.9\pm0.2$}}&\multicolumn{1}{c|}{\multirow{2}{*}{-19.0$\pm$0.2}}&\multicolumn{1}{c|}{\multirow{2}{*}{-8.1$\pm$0.7}}&\multicolumn{1}{c|}{\multirow{2}{*}{-4.0$\pm$0.1}}&\multicolumn{1}{c|}{\multirow{2}{*}{-12.9$\pm$0.1}}&\multicolumn{1}{c|}{\multirow{2}{*}{-5.6$\pm$0.1}}&\multicolumn{1}{c|}{\multirow{2}{*}{$21.7\pm0.7$}}&\multicolumn{1}{c|}{\multirow{2}{*}{40-50}}& \multicolumn{1}{c|}{\multirow{2}{*}{I}} &2.57$\pm$0.17&0.56$\pm$0.28&4.79$\pm$0.84\\
&&&&&&&&&&&&&${2.14\pm0.19}^{b}$&${0.98\pm0.57}^{b}$&${5.24\pm0.19}^{b}$\\
\cline{2-16}
&Coll.\,132&242.89&-9.04&$645.8\pm1.1$&-15.7$\pm$0.9&-6.1$\pm$2.0&-3.0$\pm$0.2&-5.2$\pm$0.1&-2.1$\pm$0.1&$27.1\pm2.1$&34&\multicolumn{1}{c|}{\multirow{3}{*}{II}}&\multicolumn{1}{c|}{\multirow{3}{*}{3.09$\pm$0.24}}&\multicolumn{1}{c|}{\multirow{3}{*}{2.22$\pm$0.11}} &\multicolumn{1}{c}{\multirow{3}{*}{0.88$\pm$0.90}}\\
\cline{2-12}
&BBJ\,1&224.26&-13.76&$368.5\pm0.7$&-14.1$\pm$0.7&3.1$\pm$0.7&-2.8$\pm$0.4&-6.8$\pm$0.1&-2.8$\pm$0.1&$23.4\pm0.9$&34&\multicolumn{1}{c|}{}&\multicolumn{1}{c|}{}&\multicolumn{1}{c|}{}&\multicolumn{1}{c}{}\\
\cline{2-12}
&BBJ\,2&238.17&-10.71&$393.7\pm0.7$&-13.7$\pm$-0.4 &0.6$\pm$0.3& -2.7$\pm$0.1 &-8.0$\pm$0.1&-2.9$\pm$0.1&$21.9\pm0.4$&34&\multicolumn{1}{c|}{}&\multicolumn{1}{c|}{}&\multicolumn{1}{c|}{}&\multicolumn{1}{c}{}\\
\cline{2-16}

&BBJ\,3&260.75&-8.36&$326.6\pm0.6$&-7.4$\pm$0.1&0.4$\pm$0.7&-6.3$\pm$0.2&-9.9$\pm$0.1&-7.6$\pm$0.1&$14.2\pm0.7$&34&\multicolumn{1}{c|}{\multirow{4}{*}{III}}&\multicolumn{1}{c|}{\multirow{4}{*}{3.27$\pm$0.15}}&\multicolumn{1}{c|}{\multirow{4}{*}{2.38$\pm$0.11}} &\multicolumn{1}{c}{\multirow{4}{*}{-8.00$\pm$0.65}} \\
\cline{2-12}
&NGC\,2547&264.42&-8.60&$386.0\pm0.5$&-6.0$\pm$0.1&-0.4$\pm$0.6&-3.9$\pm$0.1&-8.2$\pm$0.1&-5.0$\pm$0.1&$13.6\pm0.6$&34&\multicolumn{1}{c|}{}&\multicolumn{1}{c|}{\multirow{6}{*}{${2.96\pm0.14}^{b}$}}&\multicolumn{1}{c|}{\multirow{6}{*}{${1.60\pm0.36}^{b}$}}&\multicolumn{1}{c}{\multirow{6}{*}{-${4.49\pm0.37}^{b}$}}\\
\cline{2-12}
&NGC\,2451B&252.31&-6.81&$363.7\pm0.6$&-9.4$\pm$0.3&0.7$\pm$0.7&-5.0$\pm$0.1&-8.9$\pm$0.1&-6.0$\pm$0.1&$16.6\pm0.8$&34&\multicolumn{1}{c|}{}&\multicolumn{1}{c|}{}&\multicolumn{1}{c|}{}&\multicolumn{1}{c}{}\\
\cline{2-12}
&Coll.\,135&248.81&-11.03&$299.4\pm0.3$&-8.4$\pm$0.2&2.3$\pm$0.4&-5.1$\pm$0.1&-10.0$\pm$0.1&-6.4$\pm$0.1&$16.1\pm0.4$&34&\multicolumn{1}{c|}{}&\multicolumn{1}{c|}{}&\multicolumn{1}{c|}{}&\multicolumn{1}{c}{}\\
\cline{2-12}
&Coll.\,140&244.95&-7.77&$383.9\pm0.8$&-11.8$\pm$0.6&-1.4$\pm$1.2&-4.6$\pm$0.2&-7.9$\pm$0.1&-5.0$\pm$0.1&$21.1\pm1.4$&34&\multicolumn{1}{c|}{}&\multicolumn{1}{c|}{}&\multicolumn{1}{c|}{}&\multicolumn{1}{c}{}\\
\cline{2-12}
&UBC\,7&248.73&-13.51&$276.3\pm0.3$&-8.3$\pm$0.2&1.4$\pm$0.5&-4.3$\pm$0.2&-10.5$\pm$0.1&-5.9$\pm$0.1&$16.9\pm0.6$&34&\multicolumn{1}{c|}{}&\multicolumn{1}{c|}{}&\multicolumn{1}{c|}{}&\multicolumn{1}{c}{}\\

\cline{2-16}
&Haffner\,13&245.05&-3.60&$560.3\pm1.5$&-23.7$\pm$0.6&-11.3$\pm$1.1&-1.5$\pm$0.4&-8.2$\pm$0.1&-2.5$\pm$0.1&$33.7\pm1.2$&34& V & - & - & -\\

\hline
\hline
\end{tabular}

%\end{adjustwidth}
%\end{center}
\tablecomments{$^a$ The observed proper motions include the effect of the solar motion.
$^b$ From \citet{Cantat-Gaudin(2019b)}}
\label{tab:sums}
\end{table*}
%\end{landscape}
\end{turnpage}
\clearpage
%}
%--

\end{CJK*}
\end{document}